\documentclass[a4paper]{article}
\pdfoutput=1 % if your are submitting a pdflatex (i.e. if you have
\usepackage{ae,aecompl}
\usepackage{jheppub} % for details on the use of the package, please
\usepackage[T1]{fontenc}
\usepackage[none]{hyphenat}
\usepackage[italian,english]{babel}
\usepackage{hyperref}
\usepackage{changepage}
\usepackage{ifpdf}
\usepackage{subfigure}
\usepackage{amssymb}
\usepackage{amsfonts}
\usepackage{epsf}
\usepackage{rotating}
\usepackage{graphicx}
\usepackage{amsmath}
\usepackage{fancyhdr}
\usepackage{lineno}
\usepackage{babel}
\usepackage{graphics}
\usepackage{pstricks}
\usepackage{color}
\usepackage{multirow}
\usepackage{slashed}
\usepackage{physics}
\usepackage{caption}
\usepackage{tabularx, array,booktabs, makecell}
\usepackage[toc,page]{appendix}
\usepackage{pifont}
\usepackage{orcidlink}  % ORCID ID 

\hypersetup{
	colorlinks=true,
	linkcolor=[rgb]{0.0, 0.2, 0.6},      % dark blue for URL
	citecolor=[rgb]{0.0, 0.5, 0.1},   % dark green for citations
	urlcolor=[rgb]{0.0, 0.2, 0.6}      % dark blue for internal links
}

\usepackage{tikz}
\usepackage{tikz-feynman}
\tikzfeynmanset{compat=1.0.0}

\newcolumntype{P}[1]{>{\centering\arraybackslash}p{#1}}
\newcolumntype{M}[1]{>{\centering\arraybackslash}m{#1}}

\title{Illuminating Degenerate Dark Sector of Inert Doublet Model at Muon Collider}

\author[a]{Anupam Ghosh\,\orcidlink{0000-0003-4163-4491}\,}
\author[b]{\!\!, Partha Konar\,\orcidlink{0000-0001-8796-1688}\,}
\author[b]{\!\!, Chandrima Sen\,\orcidlink{0000-0003-4520-1263}\,}
\author[b]{\!\!, Bhavya Thacker\,\orcidlink{0009-0007-6306-7746}\,}

\affiliation[a]{Department of Physics, Indian Institute of Technology Guwahati, North Guwahati, 781039, Assam, India}
\affiliation[b]{Theoretical Physics Division, Physical Research Laboratory, Ahmedabad, 380009, Gujarat, India}

\emailAdd{anupamg@rnd.iitg.ac.in}
\emailAdd{konar@prl.res.in}
\emailAdd{chandrima@prl.res.in}
\emailAdd{b.thacker172@gmail.com}

\abstract{
The Inert scalar Doublet Model (IDM) presents a simple yet elegant framework for a scalar dark sector where the lightest mode functions as a viable dark matter candidate under a $\mathbb{Z}_2$ symmetry. With TeV-scale particles accessible to the Large Hadron Collider (LHC) or future colliders, probing different dark matter scenarios within IDM provides an exciting opportunity. While the Higgs portal dark matter scenario is extensively well-studied at the LHC, probing the degenerate scalar dark sector presents some unique challenges, not only in detecting excessively soft decay products and the tiny production cross-section expected at the higher masses. The present study explores the potential of a forward muon facility at a future muon collider to uncover this elusive degenerate dark sector.
}

\preprint{\today}

\allowdisplaybreaks

\begin{document}
\maketitle
\flushbottom
	
%======================================================================================
\section{Introduction}
\label{intro}
%======================================================================================

The true nature of dark matter (DM) remains one of the most profound open questions in modern physics. A wide range of astrophysical and cosmological observations, including galactic rotation curves \cite{Sofue:2000jx}, gravitational lensing \cite{Clowe:2006eq}, structure formation, and cosmic microwave background (CMB) anisotropies \cite{Planck:2018nkj} provide strong evidence for a non-luminous, non-baryonic component of the universe. However, the underlying particle nature of DM remains elusive. Viable candidates span a broad spectrum, including Weakly Interacting Massive Particles (WIMPs) produced via thermal freeze-out \cite{Jungman:1995df}, an axion-like particle motivated by solutions to the strong CP problem \cite{Kim:2008hd,Peccei:1977hh}, or other well motivated theoretical scenarios \cite{Bertone:2004pz, Arcadi:2017kky, Ghosh:2022rta, Srivastava:2025oer, Ghosh:2024nkj}. Many of these models also offer testable signatures at colliders \cite{LHCDarkMatterWorkingGroup:2018ufk, Ghosh:2023xhs, Ghosh:2024boo}. Among them, the Inert Doublet Model (IDM) provides a simple and minimal extension of the Standard Model (SM) with a stable electroweak-scale DM candidate, stabilized by a discrete symmetry \cite{Barbieri:2006dq, Cirelli:2005uq}.

The IDM extends the SM scalar sector by adding a second $SU(2)_L$ scalar doublet, odd under a $\mathbb{Z}_2$ symmetry. This symmetry ensures the stability of the lightest $\mathbb{Z}_2$-odd particle, making it a natural DM candidate. Despite its minimal content, the model exhibits rich phenomenology and is constrained by relic abundance data, direct detection bounds, and collider searches \cite{Ghosh:2021noq,  Datta:2016nfz, Ghosh:2025agw, Diaz:2015pyv, Ilnicka:2015jba, Arhrib:2013ela, Dercks:2018wch, Gustafsson:2012aj, Miao:2010rg, Cao:2007rm, Kalinowski:2018kdn, Bhardwaj:2019mts,Kalinowski:2020rmb}. Primarily, two main regions of parameter space satisfy the observed DM relic density: the hierarchical regime and the compressed spectrum \cite{Goudelis:2013uca, Gustafsson:2010zz, Belyaev:2016lok, Tsai:2019eqi, Kanemura:2016sos}. In the hierarchical case, annihilation near the Higgs resonance yields the correct abundance, typically requiring the DM mass to be close to half the Higgs mass. This scenario has been well studied \cite{Belyaev:2018ext, Ghosh:2021noq, AbdusSalam:2022idz, Astros:2023gda}. 
%In contrast, the compressed spectrum features small mass splittings among the inert scalars, leading to soft visible decay products and suppressed production rates. 
In contrast, the compressed spectrum features small mass splittings among the inert scalars, resulting in soft visible decay products that remain nearly invisible at the detectors, and suppressed production rates if some of the heavier modes in the dark sector are produced in the collider. These characteristics limit current LHC sensitivity, especially for heavy DM. While some disappearing track searches have begun probing this region \cite{Belyaev:2020wok, CMS:2020atg, ATLAS:2022rme, Mahbubani:2017gjh}, much of the viable parameter space remains unexplored.

Lepton colliders, particularly the proposed multi-TeV Muon Collider (MuC) \cite{Black:2022cth, Accettura:2023ked, InternationalMuonCollider:2024jyv, MuCoL:2024oxj, InternationalMuonCollider:2025sys, AlAli:2021let}, offer a promising avenue to probe such challenging scenarios. The clean environment without much hadronic activity, absence of pile-up, and negligible QCD background make the MuC ideal for investigating weakly interacting and compressed sectors. At high energies, vector boson fusion (VBF) processes become the dominant production mode, while Drell-Yan processes are suppressed. The VBF cross section grows logarithmically with centre-of-mass energy, making the MuC particularly suited for inert scalar pair production in the degenerate mass regime \cite{Costantini:2020stv, Ruiz:2021tdt, Bandyopadhyay:2024plc}.
The characteristic VBF signal involves two forward muons and missing energy, arising from $t$-channel gauge boson exchange in processes such as $\mu^+ \mu^- \to \mu^+ \mu^- SS$, where $S$ denotes neutral or charged inert scalars. Due to the small mass splittings and large masses of the scalars, the final state has little central activity and only modest missing energy. This makes conventional search strategies based on large visible or missing energy less effective. Instead, tagging the energetic forward muons becomes essential for signal reconstruction and background suppression.

We perform a detailed scan of the compressed IDM parameter space, incorporating constraints from relic density, direct detection, and collider data. The viable region typically features heavy DM and a suppressed Higgs portal coupling that lies above the neutrino floor, making it accessible at colliders. From this parameter space, we select two representative benchmark points for collider analysis.
A comprehensive simulation of signal and background processes is carried out at a 10 TeV MuC, incorporating realistic detector assumptions. We first employ a cut-based analysis using high-level kinematic observables such as the di-muon invariant mass, missing invariant mass, angular separation between two forward muons, etc. To assess detector performance, we consider three benchmark scenarios for the muon energy resolution: $\delta_{\text{res}} = 1\%$, $5\%$, and $10\%$. The results show a marked decline in sensitivity with worsening resolution.
To further improve discrimination, we implement a multivariate analysis using the \textsc{XGBoost} algorithm. This machine learning technique captures correlations among variables that cut-based strategies cannot easily exploit. Even in low-resolution scenarios, the multivariate approach enhances the signal significance and improves the reach of the MuC for compressed spectra.

This paper is organized as follows. In \autoref{sec:model}, we present the IDM framework. \autoref{sec:constraints} outlines theoretical and experimental constraints. Dark matter phenomenology and benchmark selection are discussed in \autoref{sec:DMconstraints}. The collider analysis, including both cut-based and multivariate results, is detailed in \autoref{sec:collider}. Our conclusions are summarized in \autoref{sec:conclusion}.

%======================================================================================
\section{The Model}
\label{sec:model}
%======================================================================================
In this study, we consider an extension of the Standard Model (SM) scalar sector by introducing an additional $SU(2)_L$ scalar doublet, denoted by $\Phi_2$, with hypercharge $Y=1/2$. To stabilize the lightest neutral scalar and suppress flavor-changing neutral currents (FCNC), we impose an exact, unbroken discrete $\mathbb{Z}_2$ symmetry under which $\Phi_2$ is odd, while all SM fields are even. This discrete symmetry ensures that the new scalars do not acquire vacuum expectation values (VEVs) and prohibits their couplings to SM fermions. As a result, the new neutral scalars do not mix with the SM Higgs boson after electroweak symmetry breaking, rendering them inert in nature. 

The two scalar doublets take the following form \cite{Branco:2011iw}:
\begin{eqnarray}
	\Phi_1 = \begin{pmatrix}
		G^+ \\ \frac{1}{\sqrt{2}}(v + h + iG^0)
	\end{pmatrix}, \quad \quad 
	\Phi_2 = \begin{pmatrix}
		H^+ \\ \frac{1}{\sqrt{2}}(H^0 + iA^0)
	\end{pmatrix},
\end{eqnarray}
where $v$ represents the VEV of the SM Higgs field, $h$ is the physical Higgs boson, and $G^0$, $G^+$ are the Goldstone bosons absorbed by the $Z$ and $W^+$ bosons, respectively. The inert doublet $\Phi_2$ contributes four new physical scalar states: a pair of charged scalars $H^\pm$, a CP-even neutral scalar $H^0$, and a CP-odd neutral scalar $A^0$.

The most general, renormalizable, and CP-conserving scalar potential respecting the SM gauge symmetry $\otimes$ $\mathbb{Z}_2$ symmetry is given by:
\begin{eqnarray}\label{eq:potential}
	V(\Phi_1, \Phi_2) &=& \mu_1^2\, \Phi_1^\dagger \Phi_1 + \mu_2^2\, \Phi_2^\dagger \Phi_2 + \frac{\lambda_1}{2} \left(\Phi_1^\dagger \Phi_1\right)^2 + \frac{\lambda_2}{2} \left(\Phi_2^\dagger \Phi_2\right)^2 + \lambda_3 \left(\Phi_1^\dagger \Phi_1\right) \left(\Phi_2^\dagger \Phi_2\right) \nonumber \\
	&+& \lambda_4 \left(\Phi_2^\dagger \Phi_1\right) \left(\Phi_1^\dagger \Phi_2\right) + \frac{\lambda_5 }{2} \left[\left(\Phi_1^\dagger \Phi_2\right)^2 + \rm{h.c.}\right],
\end{eqnarray}
where all parameters are real to avoid CP violation in the scalar sector.
After electroweak symmetry breaking, the masses of the physical scalars are expressed as:
\begin{eqnarray}
	M_h^2 &=& \lambda_1 v^2 \\
	M_{H^0/A^0}^2 &=& \mu_2^2 + \frac{1}{2} \lambda_{H/A}\, v^2  \\
	M_{H^\pm}^2 &=& \mu_2^2 + \frac{1}{2} \lambda_3 v^2,
\end{eqnarray}
with $\lambda_{H/A} = \lambda_3 + \lambda_4 \pm \lambda_5$. In our analysis, we consider $\lambda_5 < 0$, which guarantees that the CP-even scalar $H^0$ is lighter than the CP-odd scalar $A^0$. This choice ensures that $H^0$ becomes a stable and viable dark matter (DM) candidate.

The Inert Doublet Model offers a predictive dark sector framework where the dark matter candidate $H^0$ predominantly annihilates via Higgs boson and co-annihilates with $A^0$ or $H^{\pm}$ via weak gauge boson mediated channels. Its relic abundance and direct detection prospects are strongly influenced by the scalar quartic couplings, and mass splittings in the inert sector. Additionally, theoretical constraints such as perturbativity, vacuum stability, and unitarity restrict the viable parameter space, which we consistently adhere to in our study.

In this work, we focus on scenarios where the masses of $H^0$, $A^0$, and $H^\pm$ are nearly degenerate. Such regions of parameter space are strongly motivated by both dark matter relic density requirements \cite{Chakrabarty:2021kmr} and electroweak precision measurements \cite{Bandyopadhyay:2023joz}, yet remain particularly challenging to explore at hadron colliders. At the Large Hadron Collider (LHC), compressed mass spectra result in soft visible decay products and suppressed production cross-sections at higher masses \cite{Cardona:2021ebw}, thereby limiting experimental sensitivity \cite{CMS:2019zmn}. In contrast, a future multi-TeV muon collider, with its clean experimental environment, high centre-of-mass energy, and enhanced electroweak pair production rates, provides an excellent opportunity to probe these scenarios.

In particular, we consider the vector boson fusion (VBF) production mode of inert scalar pairs, which offers comparatively larger cross-sections than alternative production processes at both hadron and lepton colliders \cite{Delannoy:2013ata,Dutta:2012xe,Han:2021udl,Costantini:2020stv}. The characteristic signature of this process features two energetic forward muons, detectable in the proposed forward detector systems of the muon collider, along with substantial missing energy. To evaluate the discovery potential of this channel, we perform a detailed cut-based analysis, supplemented by a machine learning-based multivariate analysis. In the following sections, we first delineate the parameter space compatible with relic density and direct detection limits, and subsequently investigate the prospects for discovering nearly degenerate inert scalars at a future multi-TeV muon collider.

%======================================================================================
\section{Brief of Theoretical and Collider Constraints}
\label{sec:constraints}
%======================================================================================
\textbf{Perturbative unitarity:} It demands that the scattering amplitudes for $2 \to 2$ processes involving scalars and longitudinal gauge bosons remain finite and perturbatively controlled. At tree level, this translates to the requirement that the eigenvalues of the coupled-channel scalar scattering matrix satisfy $|\Lambda_i| \leq 8\pi$. In the context of the IDM, these eigenvalues are functions of the quartic couplings $\lambda_i \, (i= 1,\dots 5)$ defined in \autoref{eq:potential}, and a detailed discussion of their derivation can be found in \cite{Ginzburg:2004vp, Ginzburg:2005dt}.
	
\textbf{Vacuum Stability Condition:} For the IDM to preserve its discrete $\mathbb{Z}_2$ symmetry after electroweak symmetry breaking, it is necessary that the inert scalar doublet $\Phi_2$ does not acquire a VEV. This requirement ensures that the vacuum corresponding to the inert phase is the true global minimum of the scalar potential. At leading order, this condition translates to the following inequality \cite{Ginzburg:2010wa}:
\begin{eqnarray}
	\frac{\mu_2^2}{\sqrt{\lambda_2}} \geq \frac{\mu_1^2}{\sqrt{\lambda_1}},
\end{eqnarray}
which must be satisfied to maintain the stability of the inert vacuum configuration.

\textbf{Boundedness from below:} To ensure the scalar potential of the Inert Doublet Model remains stable at large field values, it must be bounded from below. This requirement imposes the following conditions on the quartic couplings \cite{Deshpande:1977rw, Kanemura:1993hm, Chakrabortty:2013mha}:
\begin{eqnarray}
	\lambda_{1,2} \geq 0 ~~ \text{and,} ~~ 2 \sqrt{\lambda_1 \lambda_2} + \lambda_3 + \rm{min}(0, \lambda_4 \pm \lambda_5) \geq 0.
\end{eqnarray}

\textbf{Constraints from gauge boson widths:} Given the precise measurements of the decay widths of the electroweak gauge bosons $W^\pm$ and $Z$ at colliders \cite{L3:1999znj, CDF:1989qta}, it is important to ensure that no new kinematically allowed decay modes involving the additional scalars of the IDM contribute to these widths. In particular, if decays such as $W^\pm \to H^0 H^\pm$ or $A^0 H^\pm$, and $Z \to H^0A^0$ or $H^+H^-$ were possible, they would significantly alter the total widths of these gauge bosons, contradicting experimental observations \cite{Novikov:1999af, D0:2002wga}. To avoid this, the mass parameters of the inert scalars must satisfy the following conditions:
\begin{eqnarray}
	\hspace*{-0.8 cm}
	M_{H^0} + M_{H^\pm} \geq M_{W^\pm},~~ M_{A^0} + M_{H^\pm} \geq M_{W^\pm}, ~~ M_{H^0} + M_{A^0} \geq M_{Z} ~~ \text{and,} ~~ 2 M_{H^\pm} \geq M_Z.
\end{eqnarray} 
These inequalities ensure that the aforementioned decay channels remain kinematically inaccessible.

\textbf{Constrains from the SM Higgs measurements $h \to \gamma \gamma$ and $h \to \rm{invisible}$ :}
In the Inert Doublet Model, the charged scalar $H^\pm$ contributes to the loop induced decay of the Higgs boson to a pair of photons, thereby modifying the decay rate from its SM expectation. The partial decay width for $h \to \gamma \gamma$ in the IDM framework can be expressed as \cite{Arhrib:2012ia, Krawczyk:2013jta}:
\begin{eqnarray}
	\Gamma(h \to \gamma \gamma) = \frac{G_F\, \alpha^2\, M_h^3}{128 \sqrt{2} \pi^3} \left| \mathcal{A}_{\rm SM} + \frac{\lambda_3\, v^2}{2 M_{H^\pm}^2} \mathcal{A}_0(\tau_{H^\pm}) \right|^2,
\end{eqnarray}
where, $\mathcal{A}_{\rm SM}$ denotes the SM contribution from $W^\pm$ and top quark loops, $\mathcal{A}_0$ is the scalar loop function, and $\tau_{H^\pm}= M_{h}^2/ 4 M_{H^\pm}^2$ \cite{Posch:2010hx}. 

The latest measurement of the inclusive Higgs signal strength in the di-photon channel is reported as \cite{ATLAS:2022tnm}
\begin{eqnarray}
	\mu_{\gamma \gamma} = \frac{\sigma(pp \to h \to \gamma \gamma)}{\sigma(pp \to h \to \gamma \gamma)_{\rm SM}}= 1.04^{+0.10}_{-0.09}
\end{eqnarray}

This result, being in excellent agreement with the SM expectation, imposes significant constraints on both the mass of the charged scalar $M_{H^\pm}$ and its coupling to the Higgs boson. In particular, for lower charged scalar masses, the additional contribution to the di-photon decay width becomes sizable, thereby restricting the viable parameter space of the IDM.

Additionally, if the dark matter candidate $H^0$ is lighter than half the Higgs mass, the Higgs boson can decay invisibly into a pair of $H^0$. The corresponding partial decay width is given by \cite{Baek:2014jga}
\begin{eqnarray}
	\Gamma(h \to H^0 H^0) = \frac{\lambda_H^2 v^2}{64 \pi M_h} \sqrt{1 - \frac{4 M_{H^0}^2}{M_h^2}},
\end{eqnarray}
where $\lambda_H$ is the effective coupling governing the $h H^0 H^0$ interaction.

Such invisible decay mode contributes to the total Higgs width and consequently reduce the visible branching fractions. Recent LHC measurements constrain the branching ratio for invisible Higgs decays to be less than 0.107 at the 95\% confidence level \cite{ATLAS:2023tkt}. This places bounds on the combination of $M_{H^0}$ and $\lambda_H$. In the present analysis, we focus on heavier dark sector  where this decay is kinematically forbidden, thus automatically satisfying the invisible decay constraints.

\textbf{Electroweak precision observables:} The IDM is also constrained by electroweak precision observables, particularly through the oblique parameters $S$, $T$, and $U$, which encapsulate potential new physics effects on the gauge boson propagators \cite{Peskin:1991sw}. These parameters are sensitive to mass splittings within the scalar sector, particularly between the charged scalar and the neutral components, as well as between the two neutral states. The most recent global fit yields \cite{ParticleDataGroup:2024cfk}:
\begin{eqnarray}
	S = -0.04 \pm 0.10, ~~~ T = 0.01 \pm 0.12, ~~\text{and,}~~ U = 0.05 \pm 0.09.
\end{eqnarray}
It is worth noting that the $U$ parameter remains unaffected by the addition of an inert $SU(2)_L$ doublet \cite{Chakrabarty:2021kmr}, whereas significant mass splittings can lead to noticeable shifts in the $S$ and $T$ values. In our analysis, we focus on scenarios with relatively heavy inert scalars and small mass splittings between different modes in dark sector, which naturally comply with these constraints.

\textbf{Experimental bounds from LEP and LHC:} The LEP experiment, though limited by centre-of-mass energy, has placed important lower bounds on the masses of new scalar particles through null results in searches for supersymmetric neutralinos and charginos \cite{OPAL:2002lje, OPAL:1999bfx}. These limits can be applied to the IDM since the inert scalars produce similar final states. In particular, the absence of excess events in processes like $e^+ e^- \to H^+ H^-$ and $e^+ e^- \to H^0 A^0$ implies that the charged scalar must be heavier than $\sim 70$ GeV \cite{Pierce:2007ut}. Additionally, parameter regions with light neutral scalars and small mass splittings are strongly constrained, with the allowed region requiring $M_{H^0}\geq 80$ GeV, $M_{A^0} \geq 100$ GeV for $|M_{H^0}- M_{A^0}| \leq 8$ GeV \cite{Lundstrom:2008ai}. To remain consistent with these searches, our analysis focuses on mass ranges that naturally evade these exclusions.
	
At the LHC, searches for final states featuring missing transverse energy in association with jets or leptons have been widely employed to probe supersymmetric scenarios \cite{CMS:2020bfa, CMS:2021edw, ATLAS:2022ihe, ATLAS:2024lpr}, and many of these results can be recast to constrain the IDM framework. Di-lepton plus missing energy searches \cite{ATLAS:2016vox} are particularly effective in placing bounds on the pair production of inert charged scalars when the mass difference with the neutral states is sizable. For dark matter masses above a few hundred GeV, which is the focus of our study, these constraints become significantly weaker. Further limits come from mono-jet, multi-lepton, and multi-jet plus missing energy searches conducted by the ATLAS and CMS collaborations \cite{CMS:2020bfa, CMS:2021edw, ATLAS:2022ihe, ATLAS:2024lpr, CMS:2021far, ATLAS:2020wzf}. While these channels probe certain regions of the IDM parameter space, especially for small mass splittings, their sensitivity is typically overshadowed by the stronger constraints imposed by dark matter relic density and direct detection data. These aspects will be explored in the following section. In addition, reinterpretations of invisible Higgs decay searches in vector boson fusion production modes \cite {Ngairangbam:2020ksz, Konar:2022bgc} provide additional, though generally less stringent, constraints on the model.

%======================================================================================
\section{Dark Matter Phenomenology and Parameter Space}
\label{sec:DMconstraints}
%======================================================================================
The observed abundance of dark matter in the Universe, as precisely measured by the Planck satellite, provides a crucial constraint on the parameter space of dark matter models. The latest results report a relic density of~\cite{Planck:2018vyg}
\begin{equation}
	\Omega_{\rm DM} h^2 = 0.120 \pm 0.001,
\end{equation}
which any viable dark matter candidate must either satisfy or remain under-abundant, assuming it constitutes only a fraction of the total dark matter. In the context of the IDM, the relic density is primarily determined by annihilation and co-annihilation processes involving the dark scalar components, and depends sensitively on the mass splittings between them as well as their coupling to the Standard Model Higgs. The dominant annihilation channels in the higher mass regime ($M_{H^0}\gtrsim 300$ GeV) are $H^0 H^0 \rightarrow W^+ W^-$ and $H^0 H^0 \rightarrow Z Z$, mediated by both gauge and Higgs interactions. The corresponding matrix elements for these channels involve contributions from both quartic couplings and mass splittings. The effective quartic coupling strength governing these annihilation processes can be expressed as,

\begin{eqnarray}\label{Eq:lam_eff}
	\lambda_{\rm eff} \approx \lambda_H + \frac{4 M_{H^0} \Delta m}{v^2}
\end{eqnarray}

where $\Delta m$ denotes the relevant mass splitting between heavier inert scalars and the DM, and $v \simeq 246$ GeV is the electroweak VEV.

In scenarios with small $\lambda_H$, as favored by recent direct detection constraints \cite{LZ:2024zvo, PandaX:2024qfu, XENON:2023cxc}, as we will see at the later part of this section, the annihilation cross-section is substantially reduced, particularly in the compressed mass region where the mass splittings between the dark scalars are minimal. In the degenerate mass limit, where $M_{H^0} \approx M_{A^0} \approx M_{H^\pm}$, efficient co-annihilation processes such as $H^0 A^0 \rightarrow Z \rightarrow f \bar{f}$ and $H^0 H^\pm \rightarrow W^\pm \rightarrow f \bar{f}'$ become significant. These co-annihilation channels are highly sensitive to the mass splittings, and their contributions to the total effective annihilation cross-section is also significant.

%%%  DM Relic Density  %%%
\begin{figure}[hbt]
	\begin{center}
		%\hspace*{-1.cm}
		\mbox{\subfigure[]{\includegraphics[width=0.47\linewidth,angle=0]{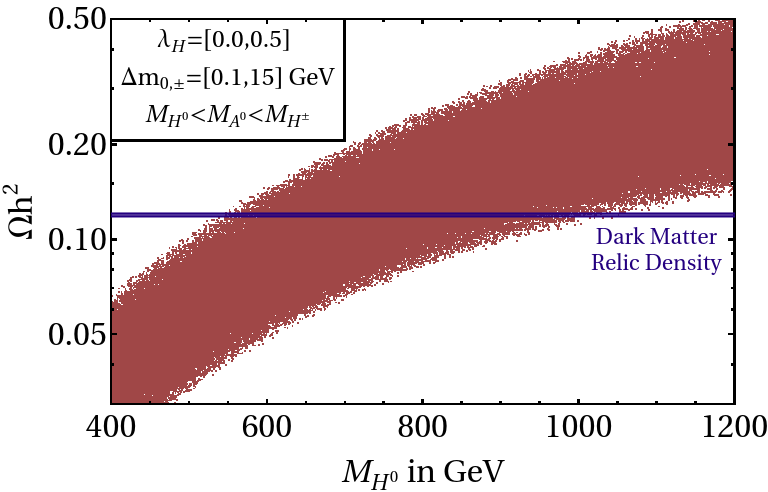}} \quad
		\subfigure[]{\includegraphics[width=0.47\linewidth,angle=0]{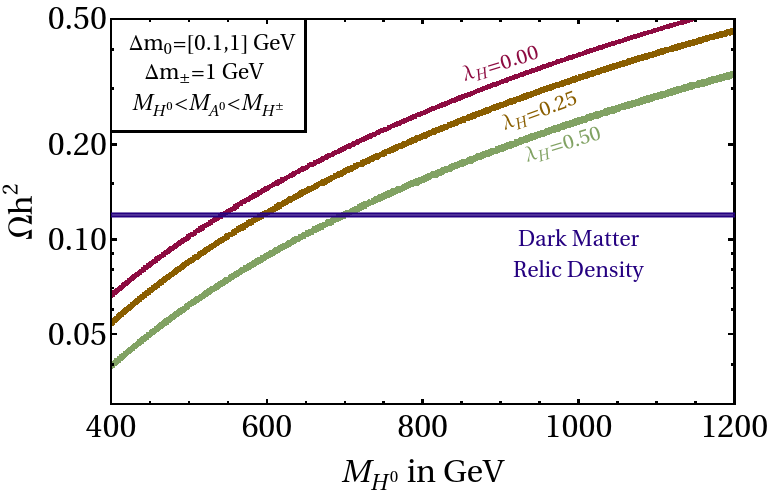}}}
		\mbox{\subfigure[]{\includegraphics[width=0.47\linewidth,angle=0]{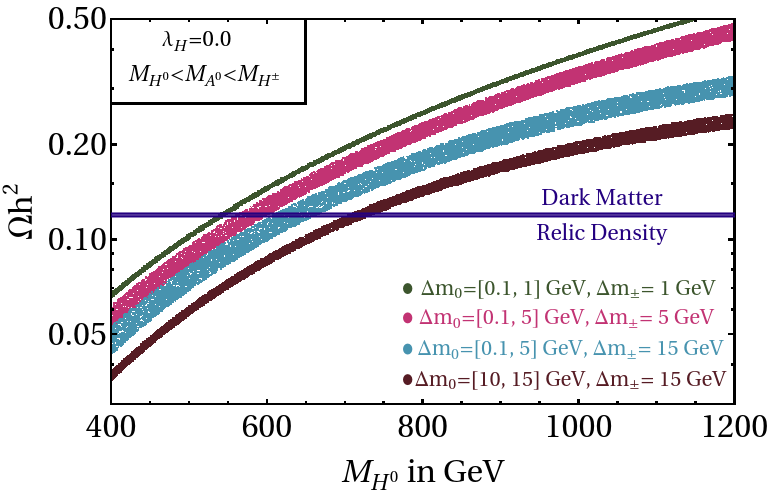}} \quad
		\subfigure[]{\includegraphics[width=0.47\linewidth,angle=0]{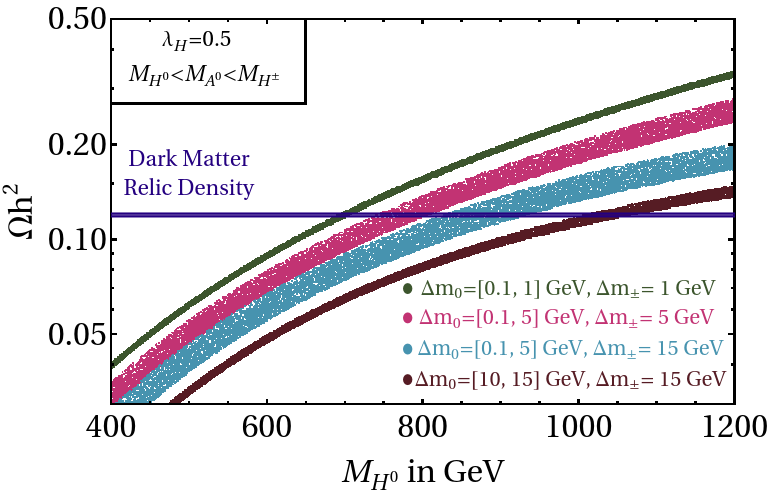}}}
		\caption{Dark matter relic density $\Omega h^2$ as a function of the dark matter mass $M_{H^0}$. Panel (a) shows the result from a random scan over the parameter space with $\lambda_H \in [0.0, 0.5]$, mass splittings among the inert scalars are varied within the range $[0.1, 15]$ GeV, and the mass hierarchy $M_{H^0} < M_{A^0} < M_{H^\pm}$ is maintained. Panels (b), (c), and (d) illustrate the relic density for specific values of $\lambda_H$ and controlled mass splittings. In (b), $\Delta m_0$ is varied within $[0.1, 1]$ GeV, $\Delta m_\pm = 1$ GeV, and $\lambda_H = 0.0,\, 0.25,\, 0.5$. Panels (c) and (d) explore different mass splitting scenarios for $\lambda_H = 0.0$ and $0.5$, respectively. The horizontal blue band, in all plots, denotes the observed dark matter relic density measured by Planck \cite{Planck:2018vyg}.}\label{fig:DMrelic}
	\end{center}
\end{figure}
%%%

For mass splittings up to $\sim$15 GeV~\footnote{The mass splitting between the charged and neutral inert scalars induced by SM loop corrections is approximately 330 MeV once the dark matter mass exceeds roughly 500 GeV \cite{Cirelli:2005uq}, and it quickly saturates at that value. Consequently, these loop effects do not significantly influence our chosen scan range of mass differences.}, co-annihilation remains efficient enough to significantly affect the relic abundance. As the mass of $H^0$ increases, the annihilation and co-annihilation cross-sections tend to decrease, resulting in a gradual rise in the relic density, as reflected in the spread of points in \autoref{fig:DMrelic} (a). Here, we present the relic density as a function of $M_{H^0}$ for a comprehensive scan over the parameter space, varying $\lambda_H$ in the range $[0.0, 0.5]$. The mass hierarchy $M_{H^0} < M_{A^0} < M_{H^\pm}$ is maintained throughout. The mass difference between the CP-odd scalar and the CP-even scalar ($\Delta m_0 = M_{A^0} - M_{H^0}$) and the mass difference between the charged scalar and the CP-even scalar ($\Delta m_\pm = M_{H^\pm} - M_{H^0}$) are varied between $0.1$ and $15$ GeV. The relic density is evaluated by solving the Boltzmann equation numerically with the help of \texttt{micrOMEGAs 6.1.15} \cite{Alguero:2023zol}. The horizontal blue band represents the observed relic density value from Planck \cite{Planck:2018vyg}. The spread of points arises from the combined effects of the parameter scan over $\lambda_H$ and mass splittings, with smaller values of $\lambda_H$ and smaller mass splittings tending to yield higher relic abundances. For the parameter scan considered, the mass of $H^0$ yielding the correct relic density extends up to about 1.1 TeV, rendering most of the TeV-scale dark matter scenarios disfavored for $\lambda_H$ values up to 0.5.

To further illustrate the combined influence of the Higgs portal coupling $\lambda_H$ and the mass splittings $\Delta m_{0,\pm}$ on the relic abundance, we present the results in \autoref{fig:DMrelic}(b), (c) and (d). Here, we systematically vary either the mass splittings while fixing $\lambda_H$, or vary $\lambda_H$ while keeping the mass splittings fixed, to disentangle their individual and combined effects. In \autoref{fig:DMrelic}(b), the relic density is shown for three representative values of the quartic coupling, $\lambda_H = 0.0,\, 0.25,\, 0.5$, while keeping the mass splitting $\Delta m_\pm$ set at 1 GeV and varying $\Delta m_0$ within the range [0.1, 1] GeV.  As expected, an increase in $\lambda_H$ enhances the annihilation cross-section through Higgs-mediated $s$-channel processes, thereby reducing the relic density for a given $M_{H^0}$.

In \autoref{fig:DMrelic}(c), we set $\lambda_H = 0.0$ and investigate the impact of varying mass splittings. Different combinations of $\Delta m_0$ and $\Delta m_\pm$ are explored within the range [0.1, 15] GeV. In this regime, the annihilation cross section is primarily governed by the mass splittings for a given $M_{H^0}$, as captured by the effective quartic coupling defined in \autoref{Eq:lam_eff}. The expression shows that $\lambda_{\text{eff}}$ increases linearly with $\Delta m$, thereby enhancing the $H^0 H^0$ annihilation cross section and reducing the resulting relic abundance. This trend is clearly reflected in \autoref{fig:DMrelic}(c), where points with larger splittings (e.g., dark brown) correspond to lower relic densities, while smaller splittings (e.g., bottle green) yield higher values of $\Omega h^2$.

\autoref{fig:DMrelic}(d) presents a similar analysis as  presented in \autoref{fig:DMrelic}(c), but for non-zero value of $\lambda_H = 0.5$. In this case, both the Higgs portal coupling and the mass splittings contribute additively to $\lambda_{\text{eff}}$ in \autoref{Eq:lam_eff}, leading to a further enhancement of the annihilation rate and consequently lower relic densities compared to the $\lambda_H = 0.0$ case. The same color scheme is used to indicate different mass splitting scenarios, with larger splittings (blue and dark brown) again associated with smaller relic abundances.

After delineating the viable parameter space from relic density constraints, it becomes essential to examine these scenarios with the limits imposed by dark matter direct detection experiments, which offer complementary and often more restrictive probes, particularly for models involving Higgs portal interactions. In the present framework, the most stringent constraints arise from the spin-independent (SI) elastic scattering of dark matter off nucleons, mediated by the Higgs boson at tree level. The corresponding cross-section is expressed as \cite{Djouadi:2011aa}
\begin{eqnarray}
	\sigma_{\rm SI} = \frac{\lambda_H^2 f^2}{4 \pi} \frac{M_n^4}{M_h^4 (M_n + M_{H^0})^2},
\end{eqnarray}
where $f$ denotes the nucleon form factor (typically varied within $0.26$–$0.63$) \cite{Mambrini:2011ik} and $M_n$ is the nucleon mass.

%%%  DM Direct Detection  %%%
\begin{figure}[hbt]
	\begin{center}
		%\hspace*{-1.cm}
		\includegraphics[width=0.75\linewidth,angle=0]{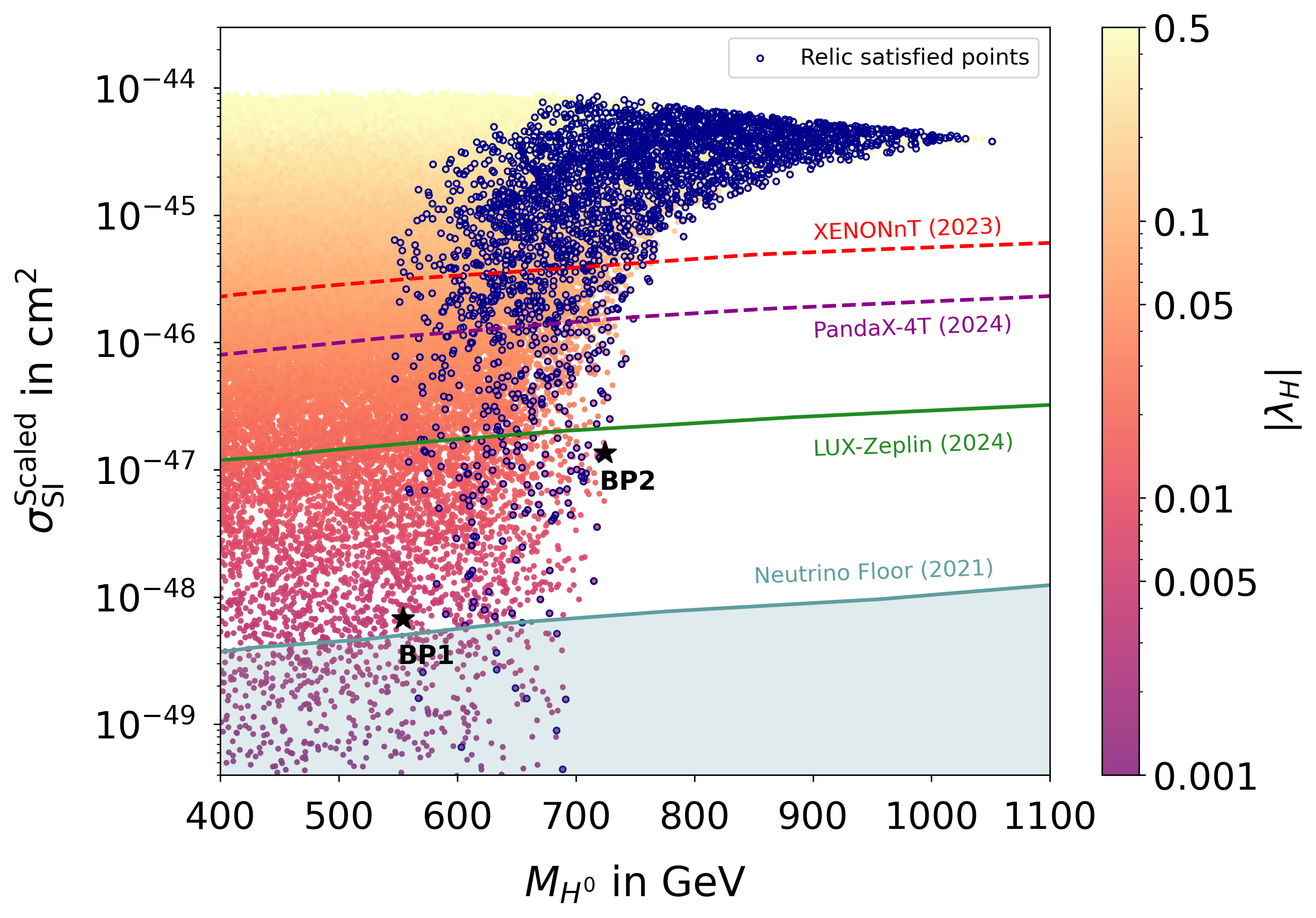}
		\caption{The scaled spin-independent cross-section as a function of dark matter mass. Current experimental limits from LZ (2024) [green solid line], PandaX-4T (2023) [magenta dashed line], and XENONnT (2023) [red dashed line] are overlaid. The neutrino floor is indicated by the light blue shaded region. A color gradient among scatter points, that are consistent with observed relic density, represents different values of $|\lambda_H|$. Blue bubbles denote points strictly satisfying the observed relic density, while the remaining points correspond to under-abundant relic scenarios.}\label{fig:DM_dirt_dec}
	\end{center}
\end{figure}
%%%

For our analysis, we compute the values of $\sigma_{\rm SI}$ for the parameter points previously shown in \autoref{fig:DMrelic}(a) that either satisfy the observed relic density or predict an under-abundance. To account for the fact that under-abundant dark matter candidates would lead to a proportionally lower scattering rate in detectors, we scale the cross-sections by a factor of $\Omega h^2 / \Omega_{\rm Planck} h^2$. The resulting scaled SI cross-sections are presented as a function of $M_{H^0}$ in \autoref{fig:DM_dirt_dec}. The color gradient of the scatter points reflects the absolute value of the Higgs portal coupling $|\lambda_H|$, with lighter shades corresponding to higher values of the coupling. Among these, the blue-circled points indicate those that yield a relic abundance consistent with the observed Planck data \cite{Planck:2018vyg}.

We overlay the most recent exclusion limits from direct detection experiments for comparison: the red dashed line represents the XENONnT (2023) \cite{XENON:2023cxc} bound, the purple dashed line corresponds to PandaX-4T (2024) \cite{PandaX:2024qfu}, and the solid green line denotes the LUX-ZEPLIN (2024) \cite{LZ:2024zvo} exclusion curve. Additionally, the projected neutrino floor is shown by the light blue shaded region at the bottom, which marks the threshold below which coherent neutrino scattering becomes an irreducible background for future experiments.

It is evident from \autoref{fig:DM_dirt_dec} that only a limited region of the parameter space is allowed by the most stringent LUX-ZEPLIN (2024) experiment while remaining above the neutrino floor. This region mostly corresponds to $|\lambda_H|$ values in the range $0.003$ to $0.02$. Points with higher $|\lambda_H|$ tend to have larger scattering cross-sections and are already excluded by recent direct detection experiments, while very small couplings push the cross-sections close to or below the neutrino floor. We have also highlighted two representative benchmark points (BPs) with black star symbols on the plot. These points satisfy the observed relic density, along with all other constraints discussed in \autoref{sec:constraints} and also stay within the allowed direct detection limits.

Before proceeding to their collider phenomenology, it is crucial to ensure that these BPs are also consistent with existing limits from indirect detection experiments. Such experiments aim to probe the annihilation or decay products of dark matter particles in astrophysical environments, typically through observations of gamma rays \cite{MAGIC:2016xys, HESS:2022ygk}, neutrinos \cite{KM3NeT:2024xca}, or cosmic rays \cite{AMS:2015tnn, PAMELA:2008gwm} originating from dark matter-dense regions such as the galactic center or dwarf spheroidal galaxies. Photons and neutrinos are particularly valuable messengers, as being electrically neutral, they can travel vast cosmic distances without significant deflection or absorption. Experiments like H.E.S.S. \cite{HESS:2022ygk} and Fermi-LAT \cite{MAGIC:2016xys} have placed important constraints on the possible annihilation cross-sections of dark matter by observing the flux of these stable particles.
In the context of the Inert Doublet Model, dark matter predominantly annihilates into electroweak gauge boson final states such as $W^+W^-$ and $ZZ$. The resulting gamma-ray signals from these channels are constrained by the aforementioned experiments. However, for the parameter space of our interest, specifically $M_{H^0} \in [400,\, 750]$ GeV and $|\lambda_H| \in [0.003,\, 0.02]$, these indirect detection bounds are comparatively weaker than those from direct detection experiments. We have verified that both BP1 and BP2 remain safely within the current indirect detection limits. Having satisfied the relic density, direct detection, and indirect detection constraints, we now turn to examining the potential collider signatures of these benchmark points.

%======================================================================================
\section{Collider Analysis: Probing Dark Sector at the Muon Collider}
\label{sec:collider}
%======================================================================================

As discussed in the previous section, the parameter space of the IDM is highly constrained for a WIMP dark matter candidate. Only a narrow region remains viable, lying below the LZ exclusion limit and above the neutrino floor. Even within this allowed region, only a limited portion satisfies the correct relic density, as indicated by the blue circular points in \autoref{fig:DM_dirt_dec}, corresponding to dark matter masses in the range of 550--750 GeV. Below this mass range, the dark matter relic density is under-abundant. In light of this, we select two benchmark points (BPs) for our collider analysis, detailed in \autoref{Tab:benchmark}. We have earlier highlighted these points with black star symbols on the plot \autoref{fig:DM_dirt_dec}. For each benchmark, we provide the masses of the BSM scalar components, the mass differences between the CP-odd and CP-even neutral scalars ($\Delta m_0$), as well as between the charged Higgs and the CP-even scalar ($\Delta m_\pm$). Additionally, the values of the Higgs portal couplings ($\lambda_H$), effective quartic couplings ($\lambda_{\rm eff}$), and the corresponding relic densities are listed.

%%%  Benchmark  Points  %%%
\begin{table}[hbt]
	\renewcommand{\arraystretch}{1.5}
	\centering
	\begin{tabular}{|c|c|c|c|c|c|c|c|c|}
		\hline 
		\makecell{Benchmark \\ Points} & \makecell{$M_{H^0}$ \\ (GeV)} & \makecell{$M_{A^0}$ \\ (GeV)} & \makecell{$M_{H^\pm}$ \\ (GeV)} & \makecell{$\Delta m_0$ \\ (GeV)} & \makecell{$\Delta m_\pm$ \\ (GeV)} & $\lambda_H$ & $\lambda_{\rm eff}$ & $\Omega h^2$ \\ \hline \hline
		BP1	& 554.06 & 555.17 & 556.28 & 1.11 & 2.22 & 0.003 & 0.043 & 0.119 \\	\hline
		BP2	& 724.59 & 738.25  & 738.32 & 13.66 & 13.73 & 0.020 & 0.674 & 0.12  \\ \hline
	\end{tabular}
	\caption{Benchmark points with masses of the heavy scalar particles in dark sector with ${H^0}$ as dark matter candidate, Higgs portal couplings ($\lambda_H$), effective quartic couplings ($\lambda_{\rm eff}$), and the DM relic density values $\Omega h^2$. Here, $\Delta m_0 = M_{A^0} - M_{H^0}$ and $\Delta m_\pm = M_{H^\pm} - M_{H^0}$ are the mass differences within dark sector, smallness of which significantly contributes in  dark matter co-annihilation.}  \label{Tab:benchmark}
\end{table}	
%%%

Given these strong constraints on the  degenerate dark sector parameter space, it is imperative to explore the collider prospects of the selected benchmark points to assess the discovery potential of the $SU(2)$ doublet inert scalars. As discussed earlier, due to unbroken $\mathbb{Z}_2$ symmetry, these BSM scalars must be produced in pairs at colliders. In the viable parameter region, however, Drell-Yan production channels are significantly suppressed. This is because the $s$-channel cross section decreases with increasing centre-of-mass energy $\sqrt{s}$ at both hadronic and leptonic colliders. Away from the resonance region, the cross section falls roughly as $1/s$ or faster. On the other hand, vector boson fusion (VBF) processes become more effective, as their cross sections grow with $\log^2(s/M_V^2)$, where $M_V$ is the mass of the exchanged vector boson. This makes VBF the dominant production mechanism at high-energy colliders. We focus on scenarios with small mass differences among the BSM scalars. In such cases, pair-produced $H^0$ particles, being stable, escape the detector without decaying, contributing to missing energy. The decay of $A^0$ into $H^0$ accompanied by soft leptons, neutrinos, or jets results in visible decay products, but are expected to be too soft to be efficiently detected. A similar situation occurs for $H^\pm$ pair production. Although one might expect a sizable missing energy signal from these events, in the case of inert scalar pair production, the transverse momenta of the produced particles largely cancel out, leaving a final state that remains clean in the central region of the detector.

In principle, this signature could be investigated at the LHC by identifying the two forward jets produced via VBF. However, the substantial impact of initial state and final state QCD radiation (ISR/FSR) at hadron colliders often leads to significant footprints in the central region, which would obscure the clean signature expected from inert scalars pair productions. To overcome this limitation, we propose to probe the Inert Doublet Model with a compressed mass spectrum at a future multi-TeV muon collider (MuC), which has recently generated considerable interest within the collider physics community \cite{Bandyopadhyay:2024plc, Bandyopadhyay:2024gyg, Ruhdorfer:2024dgz, Ruhdorfer:2023uea, Asadi:2024jiy, Li:2024joa,Han:2020pif, Capdevilla:2024bwt, Ouazghour:2024twx, Ma:2024ayr, Qian:2021ihf, Sen:2021fha, Chiesa:2021qpr, Bao:2022onq, Homiller:2022iax, Azatov:2022itm, Kwok:2023dck, Ghosh:2023xbj, Asadi:2023csb, Bhaskar:2024snl, Cheung:2025uaz, Han:2025wdy, Dehghani:2025xkd, Ghosh:2025gdx, Forslund:2022xjq, Han:2020uak}. A muon collider offers several advantages for such studies. Being a lepton collider, it provides a much cleaner environment compared to hadron colliders, with almost no QCD backgrounds, and allows the entire centre-of-mass energy to be effectively utilized for the collision. Furthermore, the proposed detector design for the MuC is expected to include dedicated forward detection capabilities \cite{InternationalMuonCollider:2025sys, MAIA:2025hzm,Capdevilla:2021fmj}, enabling the efficient identification of VBF muons ($\mu_f$), which is a crucial feature for isolating these otherwise elusive signatures.

Despite these advantages, the MuC environment presents its own challenges. The most notable being beam-induced backgrounds (BIB) \cite{Collamati:2021sbv, Ally:2022rgk}, which originate from muon decays in the beamline and produce a large number of soft, unwanted particles, predominantly in the forward region. To mitigate this, it has been proposed to install tungsten nozzles in the $|\eta| > 2.5$ region of the detector \cite{InternationalMuonCollider:2025sys, Capdevilla:2021fmj}. These nozzles act as shields, absorbing the majority of the BIB. Importantly, this arrangement will not hinder the detection of energetic VBF muons ($\mu_f$), as they possess sufficient energy to penetrate the tungsten nozzles and reach the forward detectors \cite{Ruhdorfer:2023uea, Ruhdorfer:2024dgz}, thereby preserving the feasibility of VBF-based searches at the MuC.

\begin{figure}[hbt]
	%\hspace*{-1.0cm}
	%\centering
	\tikzfeynmanset{every scalar@@/.style={thick, dashed}, every boson@@/.style={thick, decoration={snake,amplitude=1mm},decorate}, every plain@@/.style={thick}}
	\mbox{\subfigure[]{ 
			\begin{tikzpicture} [scale=0.8, transform shape]
				\begin{feynman}
					\vertex (a1){\large\(\mu^+\)};
					\vertex [right =4cm of a1] (a2){\large\(\mu_f^+\)};
					\vertex [below =3cm of a1] (a3){\large\(\mu^-\)};
					\vertex [right =4cm of a3] (a4){\large\(\mu_f^-\)};
					\vertex [right =1cm of a1] (d1){\tiny\(\textcolor{white}{d}\)};
					\vertex [above =0.18cm of d1] (d11){\tiny\(\textcolor{white}{d}\)};
					\vertex [below right =2.1cm of d1] (p1);
					\vertex [right =1cm of a3] (d2){\tiny\(\textcolor{white}{d}\)};
					\vertex [below =0.18cm of d2] (d22){\tiny\(\textcolor{white}{d}\)};
					\vertex [above right =2.1cm of d2] (p2);
					\vertex [above right =2.0cm of p1] (d3);
					\vertex [below =0.4cm of d3] (p3){\(H^0/A^0\)};
					\vertex [below right =2.0cm of p2] (d4);
					\vertex [above =0.4cm of d4] (p4){\(H^0/A^0\)};
					
					\diagram { (a1)--[plain](a2), (a3)--[plain](a4), (d11)--[boson, edge label' ={\large\(Z\)}, edge label](p1),(d22)--[boson,edge label ={\large\(Z\)}, edge label](p2), (p1)--[scalar](p3),(p2)--[scalar](p4) };
				\end{feynman}
		\end{tikzpicture}}
		\quad \quad
		\subfigure[]{
			\begin{tikzpicture} [scale=0.8, transform shape]
				\begin{feynman}
					\vertex (a1){\large\(\mu^+\)};
					\vertex [right =4cm of a1] (a2){\large\(\mu_f^+\)};
					\vertex [below =3cm of a1] (a3){\large\(\mu^-\)};
					\vertex [right =4cm of a3] (a4){\large\(\mu_f^-\)};
					\vertex [right =1cm of a1] (d1){\tiny\(\textcolor{white}{d}\)};
					\vertex [above =0.18cm of d1] (d11){\tiny\(\textcolor{white}{d}\)};
					\vertex [below right =2.1cm of d1] (p1);
					\vertex [right =1cm of a3] (d2){\tiny\(\textcolor{white}{d}\)};
					\vertex [below =0.18cm of d2] (d22){\tiny\(\textcolor{white}{d}\)};
					\vertex [above right =2.1cm of d2] (p2);
					\vertex [above right =2.0cm of p1] (d3);
					\vertex [below =0.4cm of d3] (p3){\(H^{+}\)};
					\vertex [below right =2.0cm of p2] (d4);
					\vertex [above =0.4cm of d4] (p4){\(H^{-}\)};
					
					\diagram { (a1)--[plain](a2), (a3)--[plain](a4), (d11)--[boson, edge label' ={\large\(Z/\gamma/Z\)}, edge label](p1),(d22)--[boson,edge label ={\large\(Z/\gamma/\gamma\)}, edge label](p2), (p1)--[scalar](p3),(p2)--[scalar](p4) };
				\end{feynman}
		\end{tikzpicture}}
		\quad \quad
		\subfigure[]{
			\begin{tikzpicture} [scale=0.8, transform shape]
				\begin{feynman}
					\vertex (a1){\large\(\mu^+\)};
					\vertex [right =4cm of a1] (a2){\large\(\mu_f^+\)};
					\vertex [below =3cm of a1] (a3){\large\(\mu^-\)};
					\vertex [right =4cm of a3] (a4){\large\(\mu_f^-\)};
					\vertex [right =1.7cm of a1] (d1){\tiny\(\textcolor{white}{d}\)};
					\vertex [above =0.18cm of d1] (d11){\tiny\(\textcolor{white}{d}\)};
					\vertex [below =1.5cm of d1] (p1);
					\vertex [right =1.7cm of a3] (d2){\tiny\(\textcolor{white}{d}\)};
					\vertex [below =0.18cm of d2] (d22){\tiny\(\textcolor{white}{d}\)};
					\vertex [above =1cm of d2] (p2);
					\vertex [right =1.cm of p1] (d3);
					\vertex [above right =0.6cm of d3] (s1){\(H^0/A^0/H^+\)};
					\vertex [below right =0.6cm of d3] (s2){\(H^0/A^0/H^-\)};
					
					\diagram { (a1)--[plain](a2), (a3)--[plain](a4), (d11)--[boson,edge label' ={\(Z\)}, edge label](p1),(p1)--[boson,edge label' ={\(Z\)}, edge label](d22), (p1)--[scalar, edge label ={\large\(h\)}, edge label](d3), (d3)--[scalar](s1), (d3)--[scalar](s2)};
				\end{feynman}
	\end{tikzpicture}}}
	\centering	
	\mbox{\subfigure[]{
			\begin{tikzpicture} [scale=0.8, transform shape]
				\begin{feynman}
					\vertex (a1){\large\(\mu^+\)};
					\vertex [right =4cm of a1] (a2){\large\(\mu_f^+\)};
					\vertex [below =3cm of a1] (a3){\large\(\mu^-\)};
					\vertex [right =4cm of a3] (a4){\large\(\mu_f^-\)};
					\vertex [right =2cm of a1] (d1){\tiny\(\textcolor{white}{d}\)};
					\vertex [above =0.18cm of d1] (d11){\tiny\(\textcolor{white}{d}\)};
					\vertex [below =1cm of d1] (p1);
					\vertex [below =1cm of p1] (p11);
					\vertex [right =2cm of a3] (d2){\tiny\(\textcolor{white}{d}\)};
					\vertex [below =0.18cm of d2] (d22){\tiny\(\textcolor{white}{d}\)};
					\vertex [above =1cm of d2] (p2);
					\vertex [right =1.cm of p1] (d3){\large\(H^0/A^0\)};
					\vertex [right =1.cm of p2] (d4){\large\(H^0/A^0\)};	
					
					\diagram { (a1)--[plain](a2), (a3)--[plain](a4), (d11)--[boson,edge label' ={\(Z\)}, edge label](p1),(p1)--[scalar,edge label' ={\(A^0/H^0\)}, edge label](p11),(d22)--[boson,edge label ={\(Z\)}, edge label](p2), (p1)--[scalar](d3),(p2)--[scalar](d4) };
				\end{feynman}
		\end{tikzpicture}}
		\quad \quad
		\subfigure[]{
			\begin{tikzpicture} [scale=0.8, transform shape]
				\begin{feynman}
					\vertex (a1){\large\(\mu^+\)};
					\vertex [right =4cm of a1] (a2){\large\(\mu_f^+\)};
					\vertex [below =3cm of a1] (a3){\large\(\mu^-\)};
					\vertex [right =4cm of a3] (a4){\large\(\mu_f^-\)};
					\vertex [right =2cm of a1] (d1){\tiny\(\textcolor{white}{d}\)};
					\vertex [above =0.18cm of d1] (d11){\tiny\(\textcolor{white}{d}\)};
					\vertex [below =1cm of d1] (p1);
					\vertex [below =1cm of p1] (p11);
					\vertex [right =2cm of a3] (d2){\tiny\(\textcolor{white}{d}\)};
					\vertex [below =0.18cm of d2] (d22){\tiny\(\textcolor{white}{d}\)};
					\vertex [above =1cm of d2] (p2);
					\vertex [right =1.cm of p1] (d3){\large\(H^+\)};
					\vertex [right =1.cm of p2] (d4){\large\(H^-\)};
					
					\diagram { (a1)--[plain](a2), (a3)--[plain](a4), (d11)--[boson,edge label' ={\(Z/\gamma\)}, edge label](p1),(p1)--[scalar,edge label' ={\(H^\mp\)}, edge label](p11),(d22)--[boson,edge label ={\(Z/\gamma\)}, edge label](p2), (p1)--[scalar](d3),(p2)--[scalar](d4) };
				\end{feynman}
	\end{tikzpicture}}}	
	
	\caption{The dominant Feynman diagrams for the VBF production processes: $\mu^+ \mu^- \to \mu_f^+ \mu_f^- S \bar{S}$, where, $S = H^0, A^0, H^\pm$, and $\mu_f^{\pm}$ represents the forward muons. (a), (b) denote the four-point vertex interactions, (c) illustrates the Higgs boson mediated s-channel contribution ($\gamma \gamma h, \, Z \gamma h$ vertices are neglected as it appear from the higher order corrections), and (d), (e) depict the heavy scalar mediated t-channel diagrams.}
	\label{fig:Feyn_sig}
	
\end{figure}
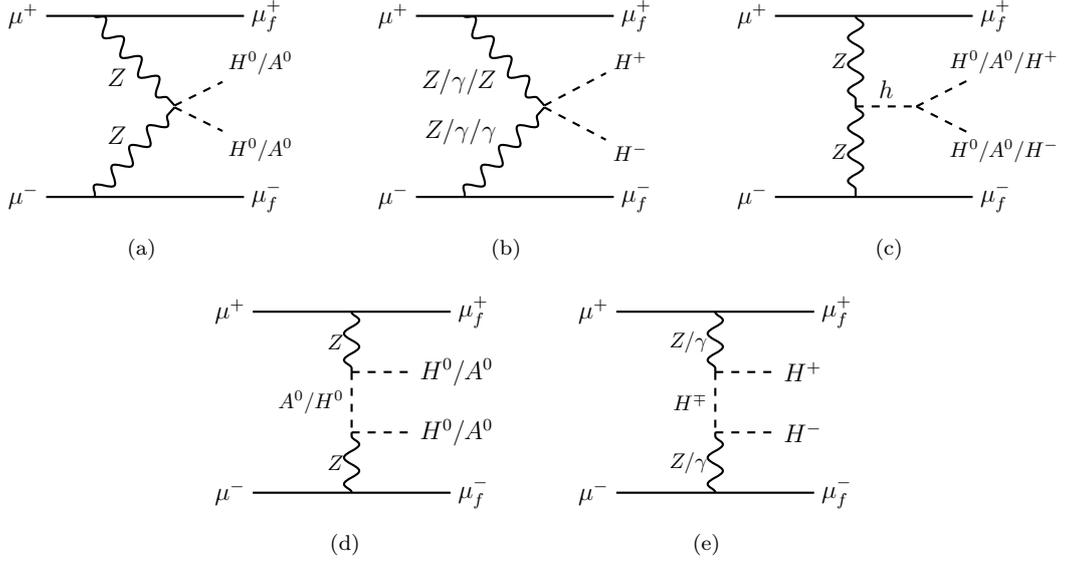

Building upon this, we propose to exploit these two forward muons as triggers in our analysis to probe the IDM in the compressed mass spectrum regime. The characteristic production processes for inert scalars via VBF are depicted in \autoref{fig:Feyn_sig}.  Diagrams (a) and (b) correspond to four-point contact interactions that lead to the pair production of neutral ($H^0, A^0$) and charged ($H^\pm$) scalars through $ZZ$ and $\gamma\gamma$ fusion processes, respectively. In the case of charged scalar production, three VBF channels get contributions from $ZZ$, $\gamma\gamma$, and $\gamma Z$ fusions whereas neutral scalar production arises solely from $ZZ$ fusion. The relevant three-point ($VSS$) and four-point ($VVSS$) interaction strengths are summarized in \autoref{tab:verfact}, where $V$ represents a $Z$ boson or a photon, and $S$ stands for the inert scalars. While photon fusion benefits from the electromagnetic coupling and hence enhances charged scalar pair production, it is noteworthy from \autoref{tab:verfact} that the $ZZSS$ vertex for neutral scalars can have a larger interaction factor.  However, the total production cross-section for charged scalars remains higher due to the combined contribution of all three channels.

Additional diagrams in \autoref{fig:Feyn_sig}(c), (d), and (e) represent $s$-channel Higgs-mediated and $t$-channel scalar-exchange contributions. The interference between these diagrams plays a crucial role in shaping the total cross-section. For neutral scalar production, a relative phase between the four-point contact interaction and the $t$-channel scalar exchange diagrams leads to destructive interference, significantly suppressing the production rate. In contrast, for charged scalar production, no such relative phase exists between the corresponding diagrams, resulting in a comparatively larger production cross-section for $H^\pm$ pairs, despite all interactions being purely gauge-mediated. To illustrate this, at a $\sqrt{s}=10$ TeV MuC, for BP1, the leading-order cross-section for the process $\mu^+ \mu^- \to \mu_f^+ \mu_f^- H^+ H^-$ is approximately $5.32 \times 10^{-2}$ fb, while the corresponding cross-section for $\mu^+ \mu^- \to \mu_f^+ \mu_f^- H^0 H^0/ A^0 A^0$ is around $7.18 \times 10^{-4}$ fb. Here, we chose $\mu_f$ in place of $\mu$ to indicate that this muon would be detected using the forward muon facility, whose definition and selection criteria are provided in \autoref{subsec:SimDet}. This quantitative difference highlights the impact of interference effects in neutral scalar production channels and demonstrates the relative advantage of charged scalar searches in such collider environments. It is also worth mentioning that the Higgs-mediated contribution shown in \autoref{fig:Feyn_sig}(c) plays a marginal role in these processes, as the $h S \bar{S}$ couplings are tightly constrained by both direct detection limits and $h \rightarrow \gamma \gamma$ decay width measurement. As a result, this collider search strategy remains effectively model-independent, relying solely on gauge couplings without sensitivity to any undetermined BSM parameters.

%%% Table for 3-point anf 4-point vertex factor %%%
\begin{table}[t]
	\begin{center}	
		%\hspace*{-1.2 cm}
		\renewcommand{\arraystretch}{1.8}
		\begin{tabular}{ |c|c|c| }
			\cline{2-3}					
			\multicolumn{1}{c|}{} &Vertex & Interaction factor   \\
			\hline
			{\multirow{6}{*}{ 3-point vertices }} & $Z_{\sigma} H^0 A^0 $ & $\frac{e}{2} \,\frac{(p_1 -p_2)_\sigma}{\cos \vartheta_w \sin \vartheta_w}$  \\
			\cline{2-3}
			& $Z_{\sigma} H^+ H^-$ & $i e (p_1 - p_2)_\sigma \, \frac{\cos(2 \vartheta_w)}{2 \cos \vartheta_w \sin \vartheta_w}$ \\
			\cline{2-3}	
			& $\gamma_{\sigma} H^+ H^-$ & $i e (p_1 - p_2)_\sigma$ \\ 
			\cline{2-3}
			& $h S \bar{S}$ & $ -i \Lambda v 
			\begin{cases} 
				\Lambda = \lambda_H, & \text{if } S = H^0 \\
				\Lambda = \lambda_A, & \text{if } S = A^0 \\
				\Lambda = \lambda_3, & \text{if } S = H^\pm 
			\end{cases} $ \\ 
			\hline
			{\multirow{4}{*}{ 4-point vertices }}& $Z_{\sigma} Z_\delta H^0 H^0 (A^0 A^0)$ & $ i e^2 \, \frac{g_{\sigma \delta}}{2 \cos^2 \vartheta_w \sin^2 \vartheta_w}$ \\ 
			\cline{2-3}
			& $Z_{\sigma} Z_\delta H^+ H^-$ & $ i e^2 \frac{g_{\sigma \delta} \cos^2 (2 \vartheta_w)}{2 \cos^2 \vartheta_w \sin^2 \vartheta_w}$ \\ 
			\cline{2-3}
			& $\gamma_{\sigma} \gamma_\delta H^+ H^-$ & $ 2i e^2 \, g_{\sigma \delta}$ \\ 
			\cline{2-3}
			& $\gamma_{\sigma} Z_\delta H^+ H^-$ & $ i e^2 \frac{g_{\sigma \delta} \cos (2 \vartheta_w)}{ \cos \vartheta_w \sin \vartheta_w}$ \\ 
			\hline
		\end{tabular}
		\caption{3-point and 4-point vertex factors that are useful in our VBF study. Here, $\vartheta_w$ is the electroweak mixing angle, $\sigma$ and $\delta$ are Lorentz indices, and $\lambda_{A/H} = \lambda_3 + \lambda_4 \pm \lambda_5$.}  \label{tab:verfact}
	\end{center}		
\end{table}	
%%%

In this analysis, we focus solely on tagging the two forward muons ($\mu_f$). Due to the nearly degenerate inert scalar spectrum, any leptons or jets originating from the decays of $A^0$ or $H^\pm$ are expected to be very soft, with transverse momenta typically below 10 GeV. Such soft decay products will fail to be detected in the central detector region of the MuC, effectively rendering our final state as a clean signature comprising only the two well-identified forward muons. The precise selection criteria used to define these forward muons are detailed in \autoref{sec:selection}. Before presenting that, we first outline the relevant SM backgrounds that can mimic the same final state in the following subsection.

%======================================================================================
\subsection{Background Identification} \label{sec:bkg}

At a MuC, an semi-invisible final state with only two forward muons can be mimicked by several Standard Model processes. The most significant of these backgrounds, ranked by their relative importance, are listed below:

\textbf{\underline{$\nu \bar{\nu}$ production:}} One of the dominant irreducible backgrounds for our signal arises from the process $\mu^+ \mu^- \to \mu_f^+ \mu_f^- \nu \bar{\nu}$. This final state originates from $Z$-boson fusion, $W^+W^-$ fusion, mixed $Z$-$W$ channels, and photon-induced contributions. In particular, diagrams involving $Z$ boson fusion lead to a final state with two forward muons and a pair of neutrinos, closely resembling our signal topology. Additionally, photon-initiated diagrams contribute significantly, especially at high energies, due to the enhancement from t-channel photon exchange. This background is particularly relevant because, like our signal, it can produce an invisible final state in the central detector while retaining two energetic forward muons.

\textbf{\underline{$W^+ W^-$ production:}} Another important background arises from the process $\mu^+ \mu^- \to \mu_f^+ \mu_f^- W^+ W^-$. Here, the $W$ bosons are produced via vector boson fusion along with two forward muons. Although the $W$ decays typically yield visible leptons or jets in the central region, a fraction of events can evade detection when both $W$ bosons decay leptonically and the resulting charged leptons either fall below the detection threshold (for instance, with $p_T < 10$ GeV) or escape detection due to reconstruction inefficiencies. Such events therefore contribute to the semi-invisible background category, with only the two forward muons detected, closely resembling our signal final state.

\textbf{\underline{$ZZ$ production:}} The process $\mu^+ \mu^- \to \mu_f^+ \mu_f^- Z Z$ also contributes to the background when both $Z$ bosons decay invisibly into neutrino pairs, which occurs with a combined branching ratio of approximately 4.0\%. Although the overall production cross-section for this process is lower than the other backgrounds discussed above, the resulting final state with two forward muons and large missing energy  mimics the signal topology, and is thus included among the relevant backgrounds in our study.

\textbf{\underline{Higgs production:}} The Muon Collider is often regarded as a Higgs factory, with significant Higgs production via vector boson fusion processes. However, the background contribution from processes where the Higgs decays invisibly, such as $h \to ZZ^* \to 4\nu$, is expected to be negligible due to the extremely small branching ratio predicted in the SM. Consequently, the number of background events from this channel is anticipated to be insignificant in comparison to the dominant backgrounds discussed above.

%======================================================================================
\subsection{Simulation Setup}\label{subsec:SimDet}

This subsection provides the details of the event simulation procedure that employed realistic detector level particle data in this analysis. To simulate the signal events of the Inert Doublet Model, we first implement the model in \texttt{FeynRules} \cite{Alloul:2013bka}, where the relevant Lagrangian is encoded and the Feynman rules are computed. The resulting \texttt{UFO} files are then exported for use in event generation. The signal events are produced, and the VBF production cross-sections are computed using \texttt{MadGraph5\_aMC@NLO v3.4.2} \cite{Alwall:2011uj,Alwall:2014hca}. Expected background events from different standard model processes are also generated in the same framework, with appropriate decay channels specified to match the relevant final states. In particular, for $\mu_f^+ \mu_f^- Z Z$ production, both $Z$ bosons were forced to decay into $\nu \bar{\nu}$ to mimic the semi-invisible signatures characteristic of the signal. During event generation, a minimum transverse momentum of $p_T > 50$ GeV is imposed on final-state muons, while no upper limit is applied on their pseudorapidity ($|\eta_{\mu_f}|$), allowing muons to be produced across the full detector acceptance. Additionally, a minimum angular separation of $\Delta R_{\mu_f\mu_f} > 0.4$ is required between the two muons. The parton-level events are then showered and hadronized using \texttt{Pythia8.3} \cite{Bierlich:2022pfr}. The resulting hadronized events are subsequently passed through \texttt{Delphes-3.5.0}\cite{deFavereau:2013fsa} to incorporate detector effects. We use the \texttt{delphes\_card\_MuonColliderDet.tcl} configuration, which includes provisions for forward muon detection. Jet clustering was performed using the \texttt{inclusive Valencia algorithm} \cite{Boronat:2016tgd,Boronat:2014hva}, with a jet radius parameter set to $R = 0.5$. Below, we describe the event selection strategy and the criteria used to isolate the signal from backgrounds in our collider analysis.

%======================================================================================
\subsubsection{Event Selection}\label{sec:selection}

Since the final state under consideration is entirely invisible in the central region, with only two forward muons tagged to trigger the process, a set of baseline event selection criteria ($\mathcal{C}_0$) is imposed to isolate the signal from background processes. The selection requirements are as follows:

\begin{itemize}
	\item Events are required to contain exactly two leading muons, ordered by their transverse momentum, lying in opposite hemispheres of pseudorapidity. This is ensured by imposing the condition $\eta_{\mu_{f_1}} \times \eta_{\mu_{f_2}} < 0$. These muons are referred to as VBF muons.
	
	\item To identify muons within the forward muon detectors, a strict requirement of $2.5 \leq |\eta_{\mu_f}| \leq 7.0$ is applied for both muons. According to the \texttt{Delphes} detector card configuration for the Muon Collider, such forward muons can be detected with an efficiency of 95\%.
	
	\item A lower energy threshold of $E_{\mu_f^{\pm}} > 500$ GeV is imposed to ensure that the forward muons possess sufficient energy to successfully penetrate the tungsten nozzles and reach the detector. This requirement also effectively suppresses any potential contamination from softer BIB muons.
	
	\item Finally, events are rejected if any visible objects such as jets, additional leptons, or photons are present in the central region. This ensures that no energy deposition occurs within the central detector over the threshold, consistent with the expected signature of the signal.	
\end{itemize}

In searches for semi-invisible final states relying solely on the tagging of two forward muons at a muon collider, the beam energy resolution plays a pivotal role. Since such event cannot be directly reconstructed, its properties must be inferred using the recoil mass technique, which depends critically on precise knowledge of both the initial beam energy and the measured momenta of the tagged forward muons. Any uncertainty in the beam energy propagates directly into the reconstructed recoil mass, smearing out potential signal features such as sharp peaks or kinematic edges associated with new invisible particles, like dark matter in our study. Furthermore, accurate reconstruction of missing energy and momentum is essential to distinguish the signal from SM backgrounds, such as $\mu^+ \mu^- \to \mu_f^+ \mu_f^- \nu \bar{\nu}$, which exactly mimic the signal topology. A high-resolution muon beam thus enhances the sensitivity of such searches by enabling tighter kinematic constraints, improved signal-to-background discrimination, and sharper recoil signatures, all of which are vital for uncovering new physics in the invisible sector.

In this analysis, three possible cases of the muon beam energy resolution ($\delta_{\rm res}$) are considered, ranging from an optimistic scenario of $\delta_{\rm res} = 1\%$, to a more realistic case of $\delta_{\rm res} = 10\%$, which is also the default setting in the Delphes Muon Collider card. The quantity $\delta_{\rm res}$ is defined as $\delta_{\rm res} = \Delta E / E$, where $E$ is the mean beam energy and $\Delta E$ denotes its uncertainty. An intermediate value of $\delta_{\rm res} = 5\%$ is also included to study the impact of gradual degradation in energy resolution on the signal significance. The analysis shows that with an energy resolution as good as $1\%$, the signal can be more effectively separated from the irreducible background. In contrast, with a poorer resolution of $10\%$, the smearing effect may occasionally cause the total reconstructed energy of the two forward-tagged muons to exceed the nominal centre-of-mass energy at a 10 TeV Muon Collider. However, such events are explicitly removed by including the requirement that the total visible energy from the two forward muons be less than or equal to the nominal centre-of-mass energy of 10 TeV. This completes our baseline selection cut ($\mathcal{C}_0$), which reads as follows:
\begin{center}
	\begin{eqnarray}\label{eq:BasicCut}
		\mathcal{C}_0 &:& n_{\mu_f} = 2, ~~~ \eta_{\mu_{f_1}} \times \eta_{\mu_{f_2}} < 0, ~~~ 2.5 \leq |\eta_{\mu_f}| \leq 7.0, ~~~ E_{\mu_f^{\pm}} > 500~\text{GeV}, \nonumber \\
		& & E_{\mu_{f_1}} + E_{\mu_{f_2}} \leq 10~\text{TeV}, ~~\text{and,} ~ n_{\ell^\pm} = n_j = n_\gamma = 0 ~\text{at central region.}
	\end{eqnarray}
\end{center}
 
%======================================================================================
\subsection{Cut-based Analysis} \label{sec:CBA}

\begin{figure}[h]
	\begin{center}
		\hspace*{-1.0cm}
		\mbox{\subfigure[]{\includegraphics[width=0.35\linewidth,angle=0]{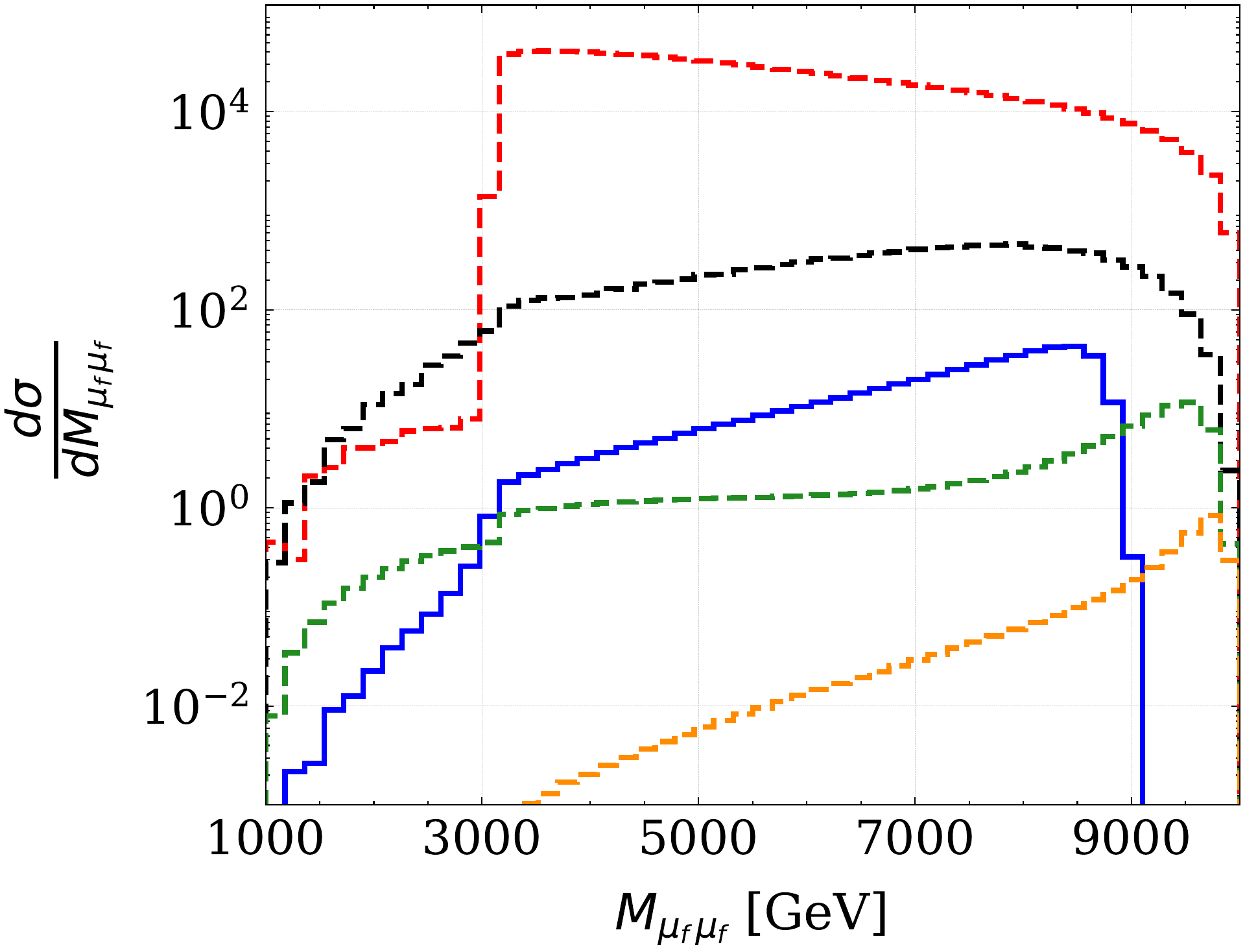}}\quad
			\subfigure[]{\includegraphics[width=0.35\linewidth,angle=0]{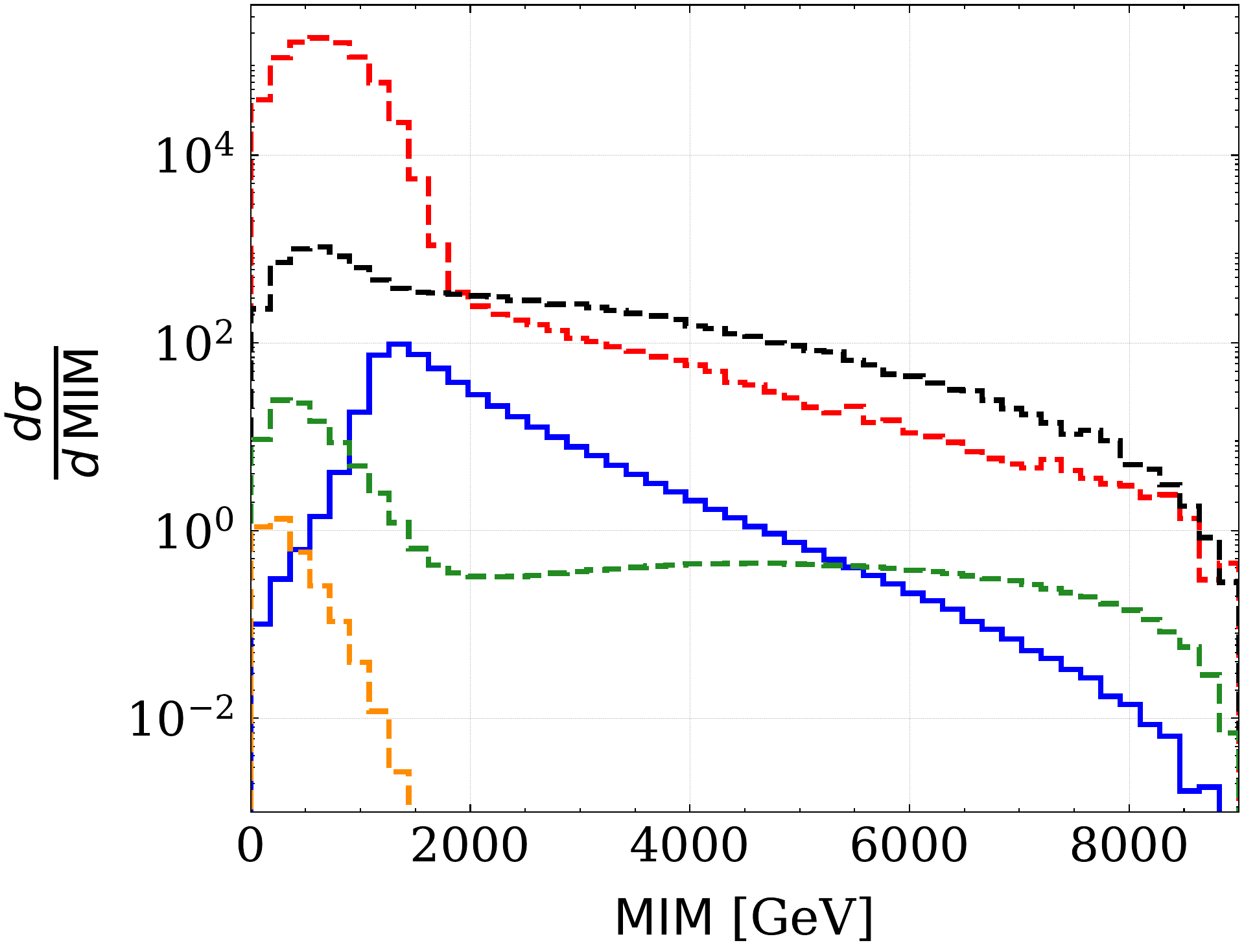}} \quad
			\subfigure[]{\includegraphics[width=0.35\linewidth,angle=0]{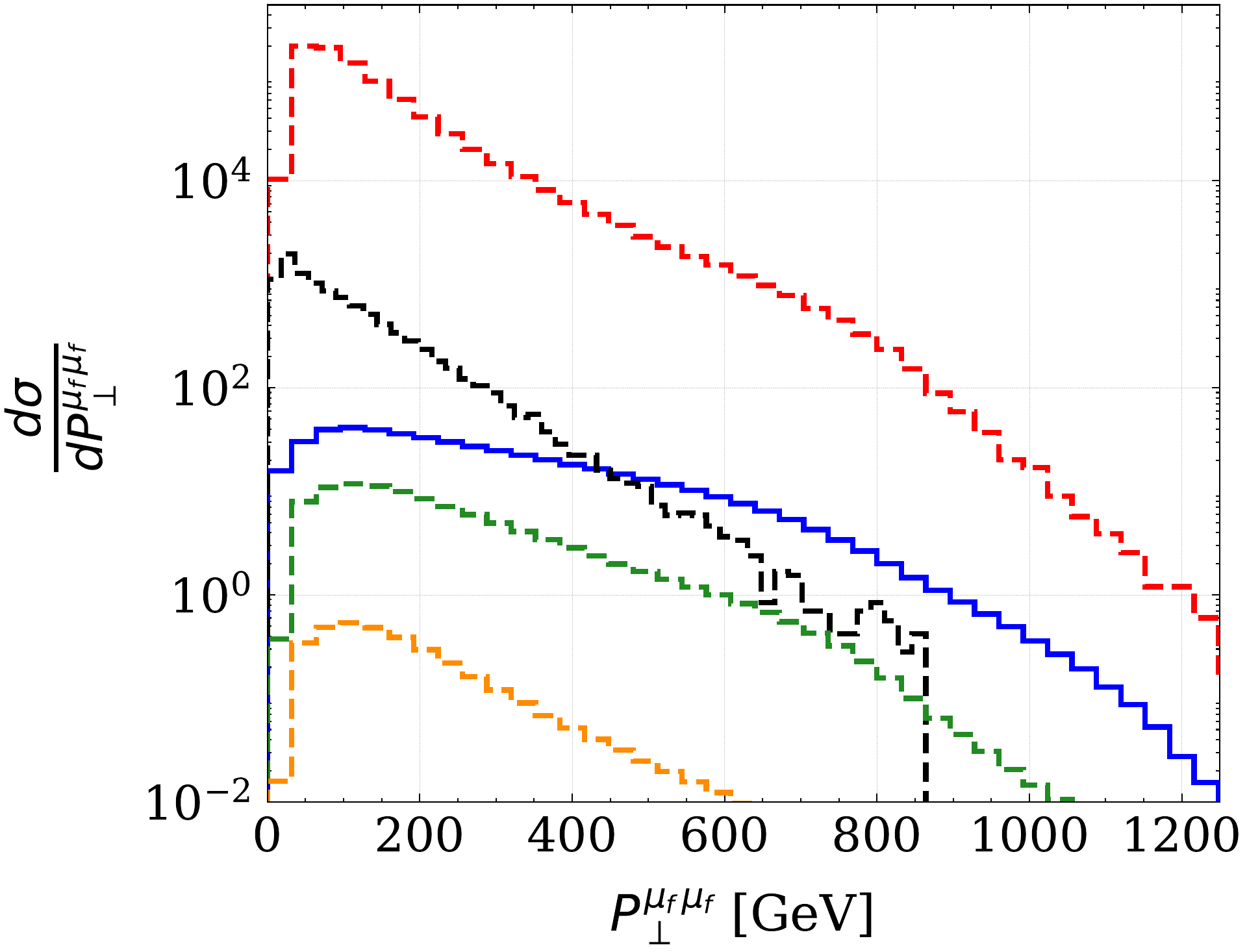}}} 	
		\hspace*{-1.0cm}		
		\mbox{\subfigure[]{\includegraphics[width=0.35\linewidth,angle=0]{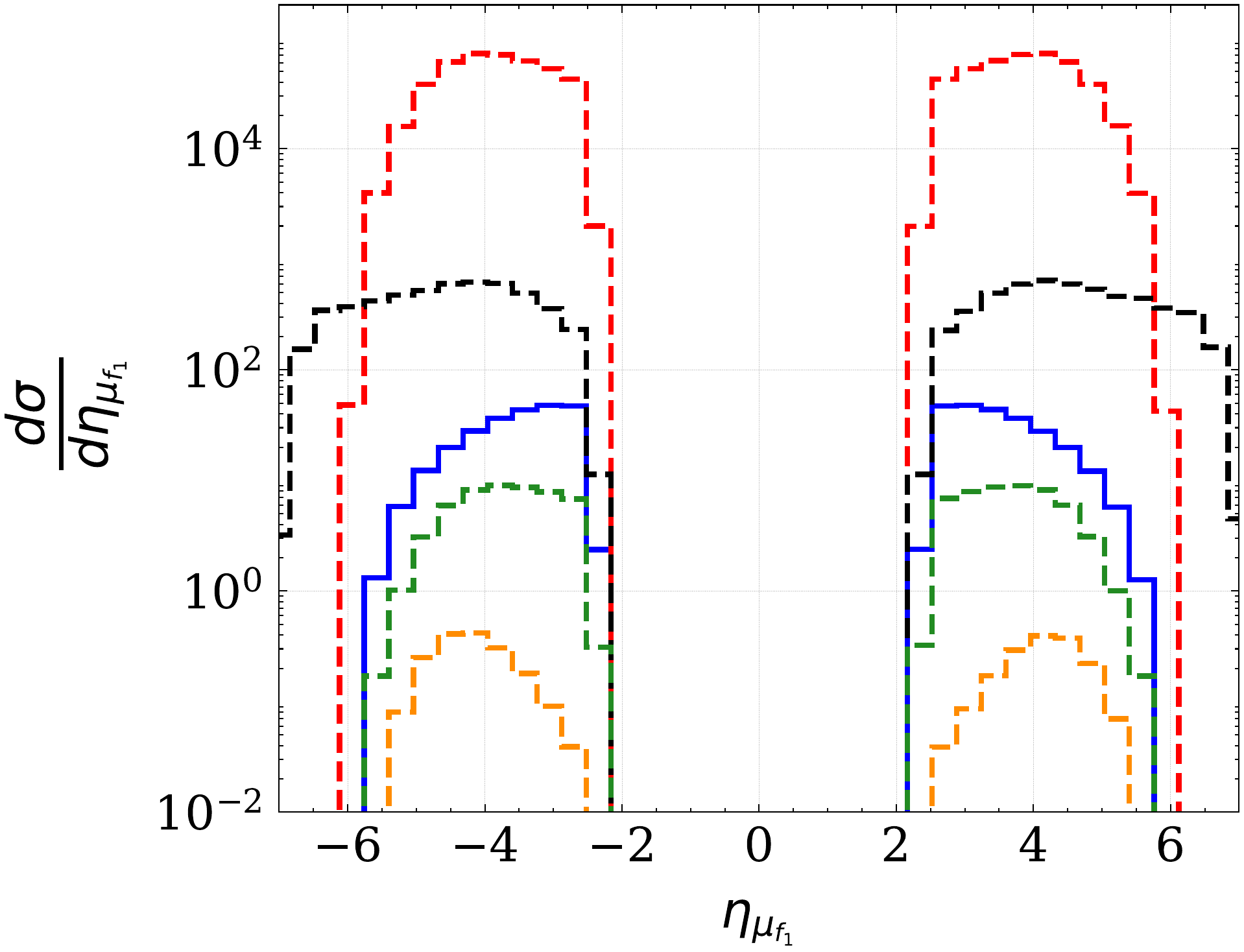}}\quad
			\subfigure[]{\includegraphics[width=0.35\linewidth,angle=0]{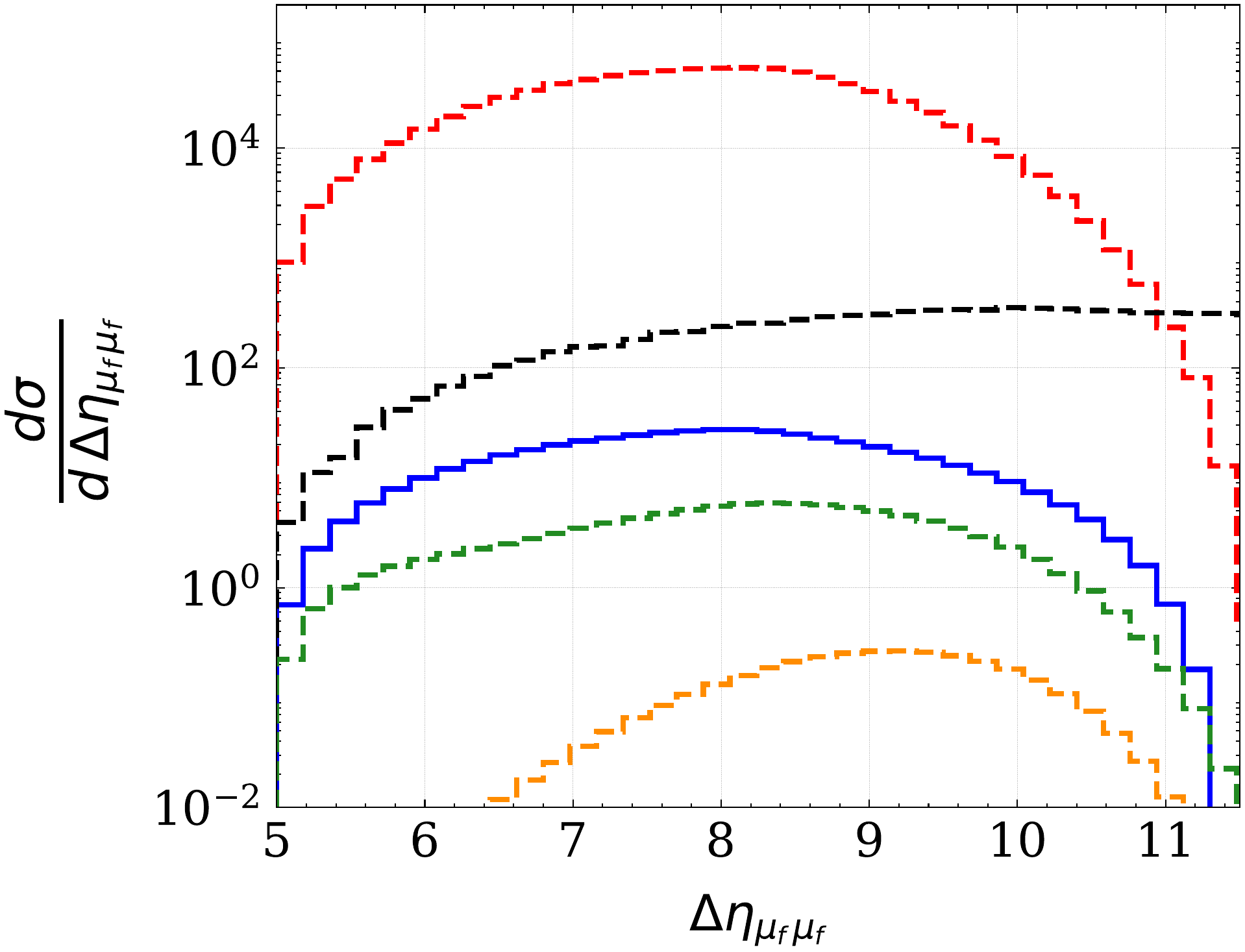}}\quad
			\subfigure[]{\includegraphics[width=0.35\linewidth,angle=0]{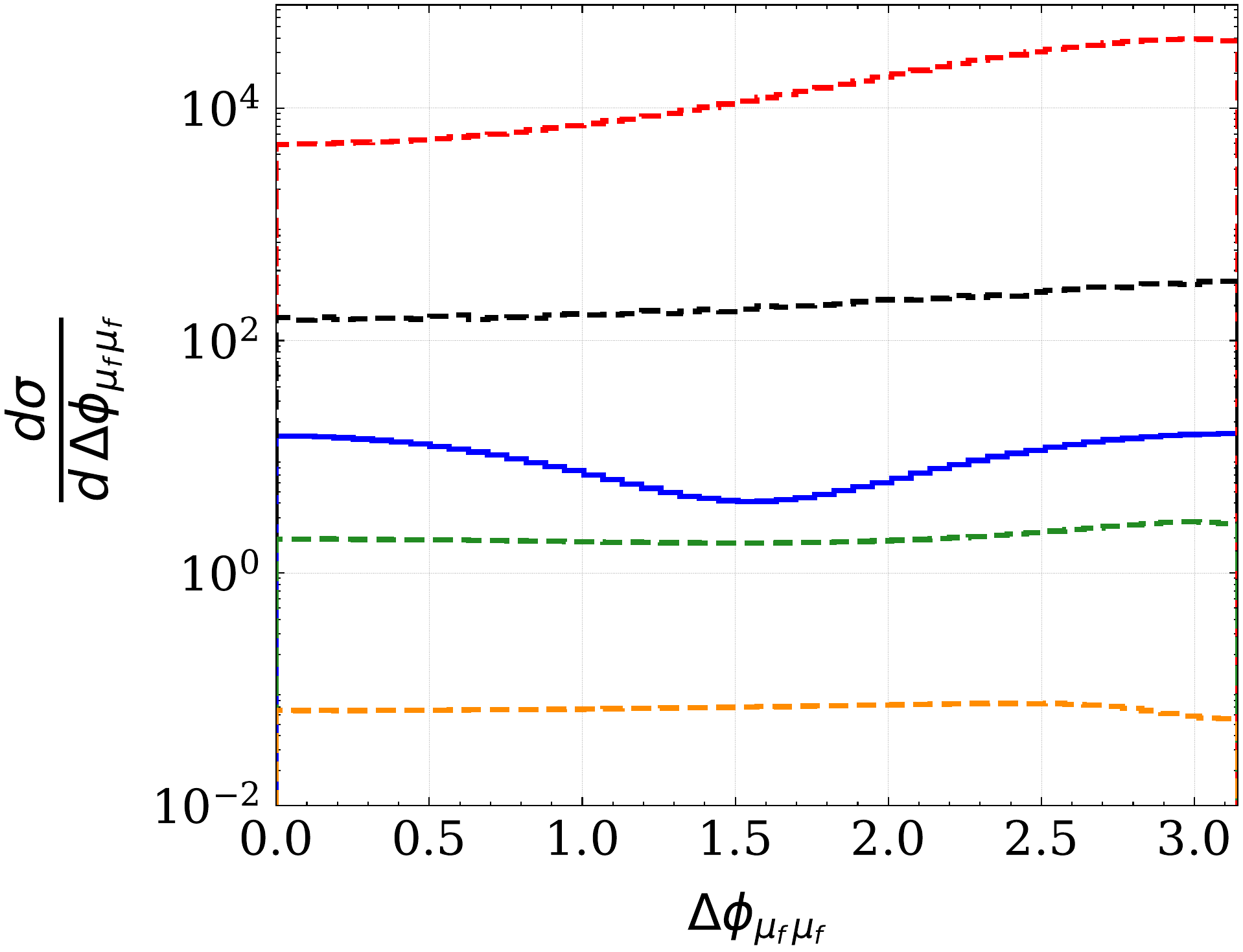}}}
		\mbox{\subfigure[]{\includegraphics[width=0.35\linewidth,angle=0]{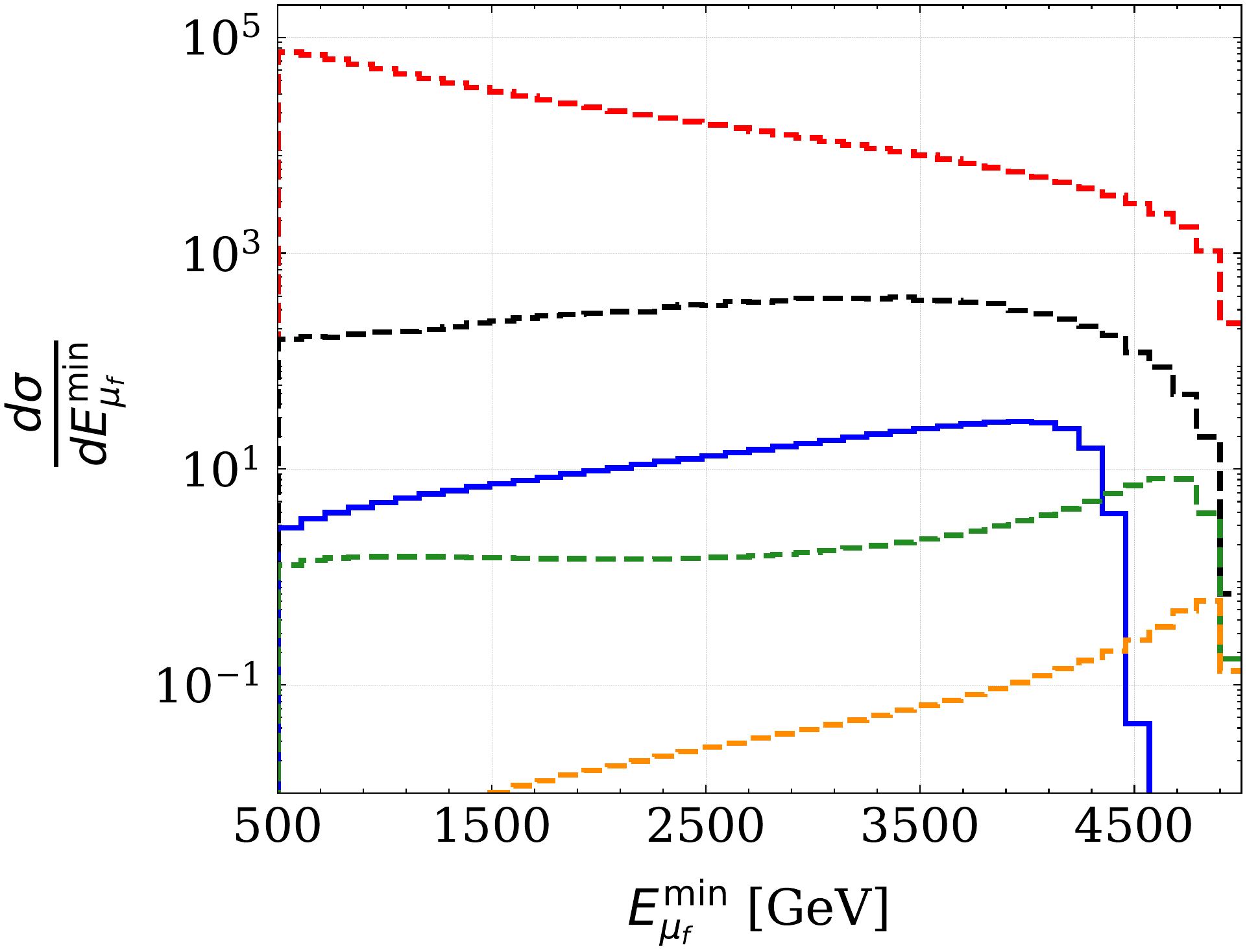}}\quad \quad
			\subfigure[]{\includegraphics[width=0.35\linewidth,angle=0]{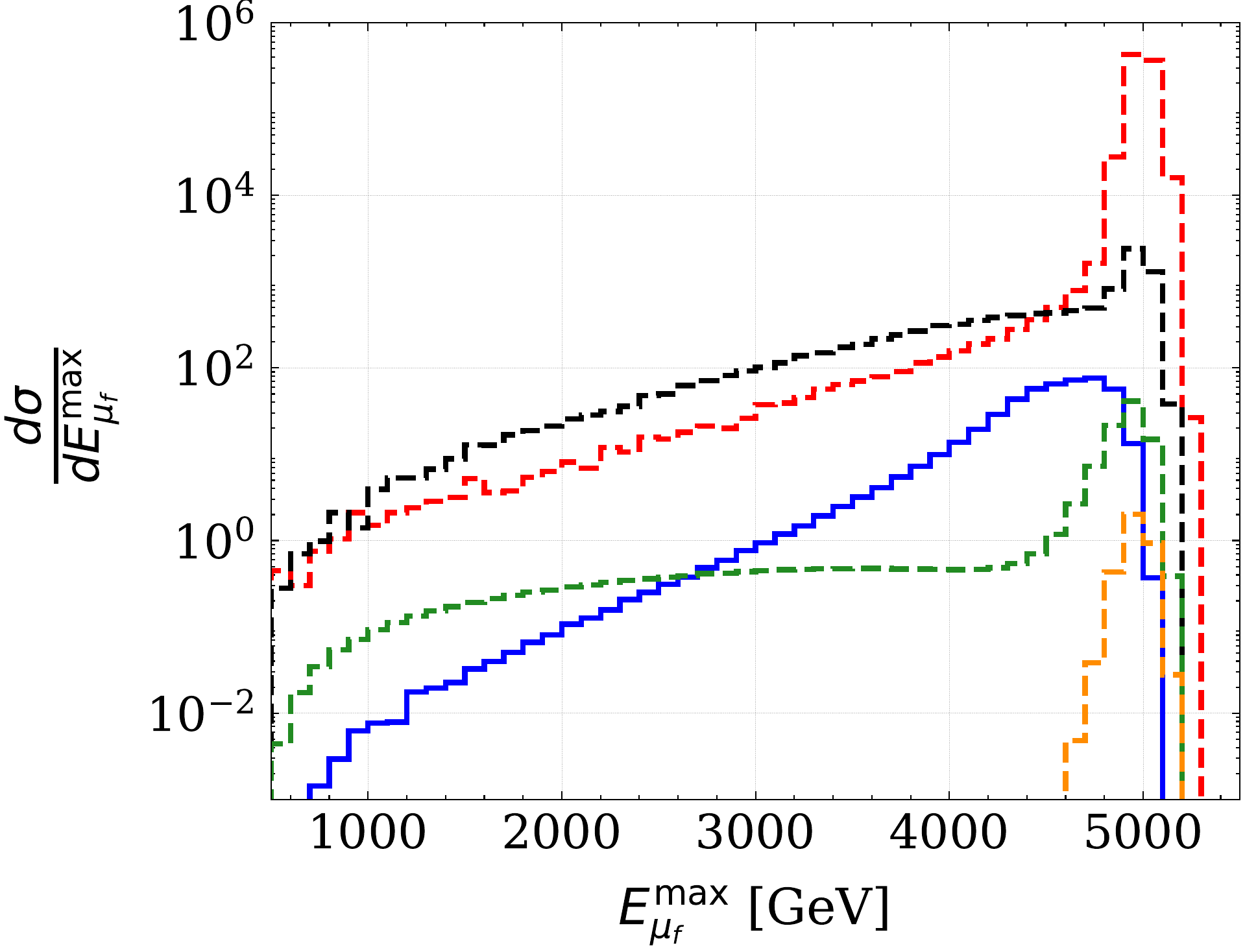}}}
		\subfigure{\includegraphics[width=0.9\linewidth,angle=0]{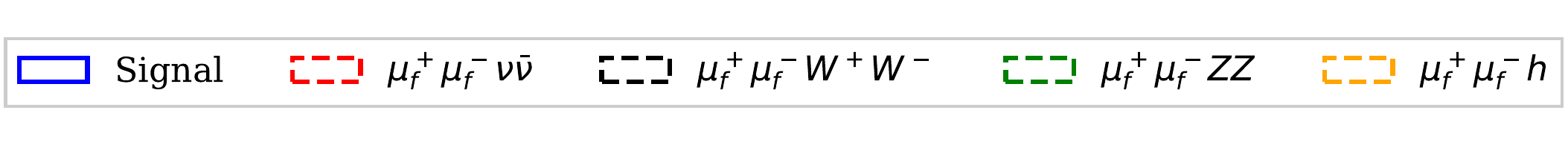}}
		\caption{Distributions of the kinematical variables for the final state $2 \mu_f$ + MET at a 10 TeV muon collider with 1\% resolution of energy. The events are weighted with respective cross-section and a luminosity of 10 ab$^{-1}$.}\label{fig:kine_dist_ER1}
	\end{center}
\end{figure}

\begin{figure}[h]
	\begin{center}
		\mbox{\subfigure[]{\includegraphics[width=0.35\linewidth,angle=0]{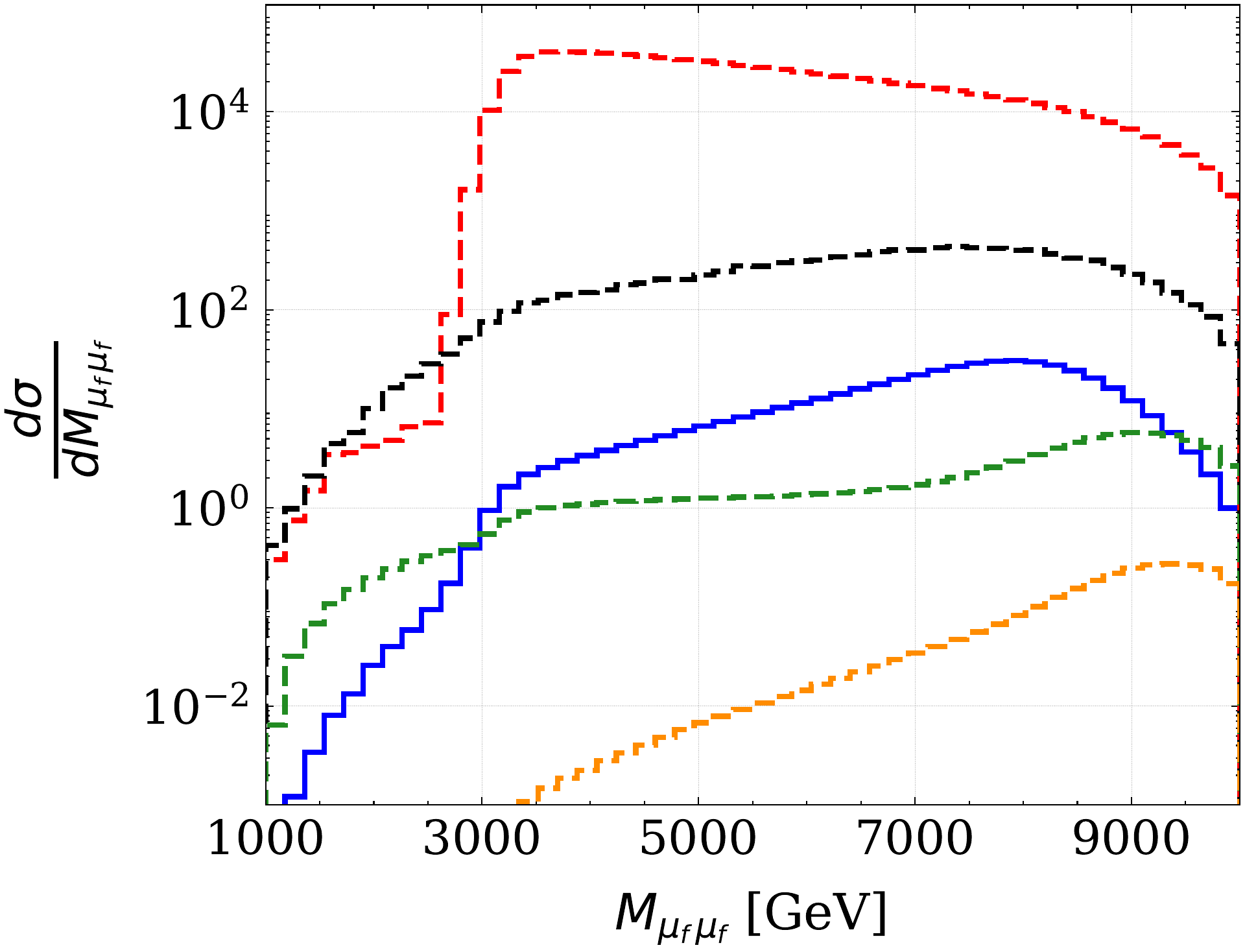}}\quad \quad
			\subfigure[]{\includegraphics[width=0.35\linewidth,angle=0]{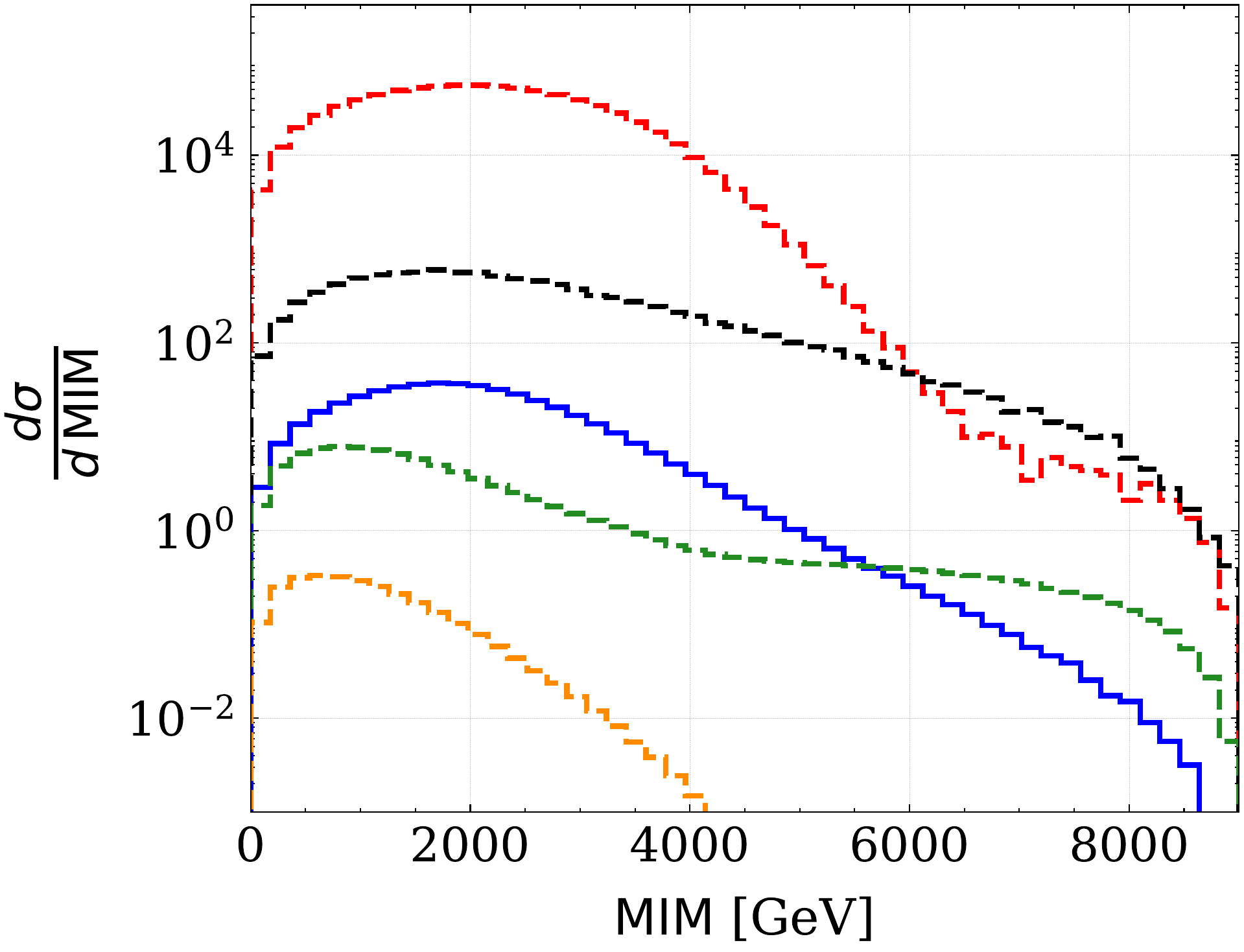}}} 	
		\mbox{\subfigure[]{\includegraphics[width=0.35\linewidth,angle=0]{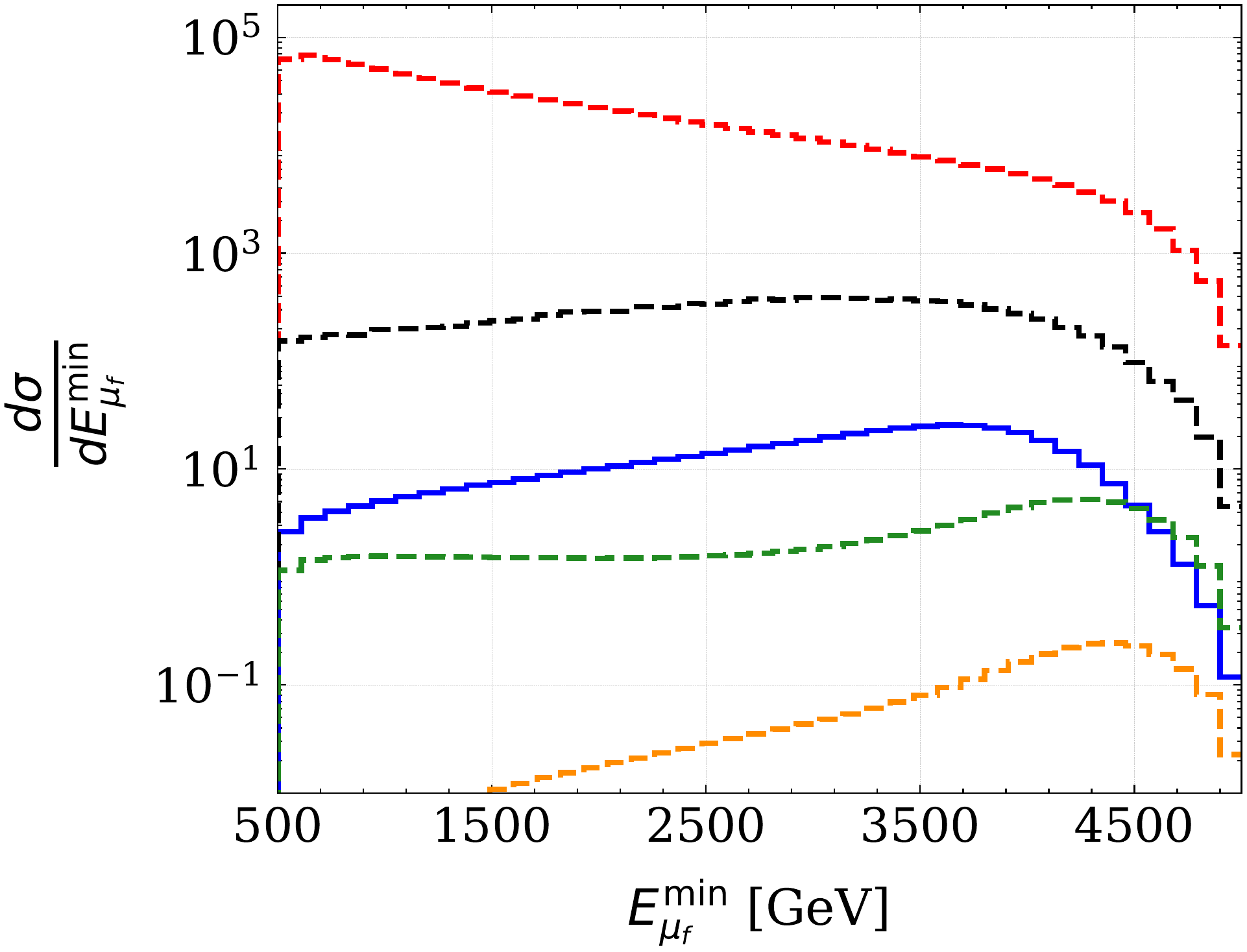}}\quad \quad
			\subfigure[]{\includegraphics[width=0.35\linewidth,angle=0]{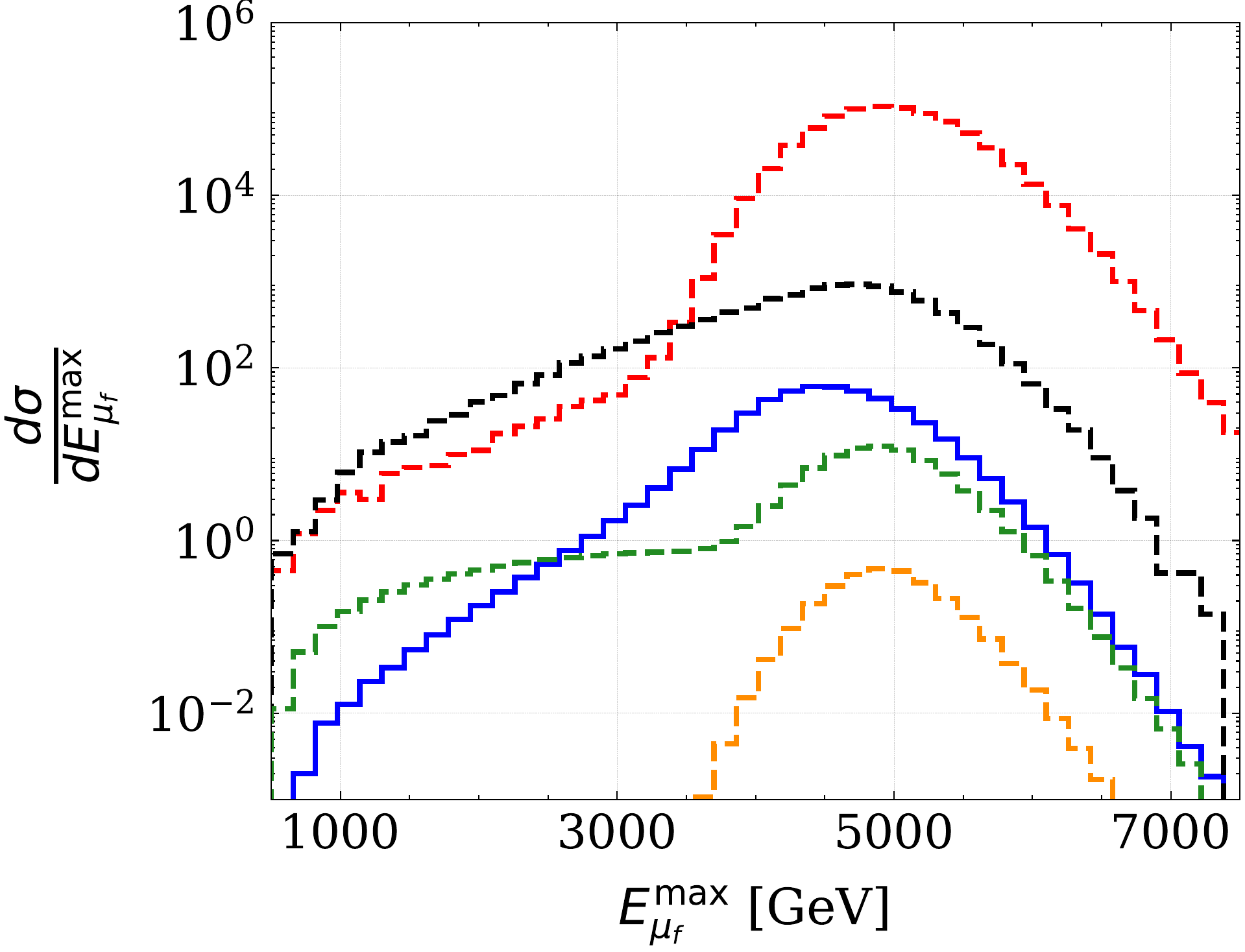}}}
		\subfigure{\includegraphics[width=0.9\linewidth,angle=0]{plots/legend.pdf}}
		\caption{Some of the discriminating distributions of the kinematical variables for the final state $2 \mu_f$ + MET at a 10 TeV muon collider with 10\% resolution of energy. The events are weighted with respective cross-section and a luminosity of 10 ab$^{-1}$.}\label{fig:kine_dist_ER10}
	\end{center}
\end{figure}

We begin with the normalized distributions of the key kinematic observables for both the signal and dominant background processes are shown in \autoref{fig:kine_dist_ER1}. These plots correspond to the BP1 scenario at a muon energy resolution of $\delta_{\rm res} = 1\%$, with all events weighted by their respective cross sections and scaled to an integrated luminosity of 10 ab$^{-1}$. To illustrate the impact of detector performance, we also present a set of corresponding distributions for an energy resolution of $\delta_{\rm res} = 10\%$ in \autoref{fig:kine_dist_ER10}. These distributions differ notably from the $\delta_{\rm res} = 1\%$ case, highlighting the deterioration in signal-background separation at poorer energy resolution, and thereby emphasizing the necessity of good beam energy resolution for reliable new physics searches. The backgrounds considered in this analysis are detailed in \autoref{sec:bkg}. In each panel, the blue histogram represents the total contribution from all signal processes, while the other colored histograms correspond to the various background components, with the color codes indicated in the plot legends.

%%%  Kinematical distribution plots (ER 1)  %%%
%%%

The first important observable is the invariant mass of the two forward muons, $M_{\mu_f\mu_f}$, which reflects the momentum transfer in the VBF process. Since, for BP1, the inert scalar mass is taken to be around 550 GeV, the minimum energy required to pair produce them is approximately 1100 GeV. Consequently, for a sufficiently good $\delta_{\rm res}$, the distribution of $M_{\mu_f\mu_f}$ for the signal should effectively die out around 8900 GeV. In contrast, in SM background processes, the forward muons can carry away nearly the entire beam energy, causing the $M_{\mu_f\mu_f}$ distribution to extend up to the centre-of-mass energy of 10 TeV, as clearly visible in \autoref{fig:kine_dist_ER1}(a). This separation worsens when the muon energy resolution degrades, as evident in the case of $\delta_{\rm res} = 10\%$. For comparison, the $M_{\mu_f\mu_f}$ distribution for this scenario is shown in \autoref{fig:kine_dist_ER10}(a), where it is observed that the signal distribution extends beyond 9.8 TeV, resulting in poor separation from the background distributions.

Another crucial discriminant is the missing invariant mass (MIM), which effectively reconstructs the invisible mass recoiling against the di-muon system. It is defined as:
\begin{eqnarray}
	\text{MIM} = \sqrt{|(2E_{\rm beam}, \mathbf{0})- p_{\mu_f^+}- p_{\mu_f^-}|^2},
\end{eqnarray}
where $E_{\text{beam}}$ is the nominal beam energy, and $p_{\mu_f^\pm}$ are the four-momenta of the forward-tagged muons. At a muon collider, where $\sqrt{s} = 2E_{\text{beam}}$, this expression simplifies to:
\begin{eqnarray}
	\text{MIM} = \sqrt{s + M_{\mu_f \mu_f}^2 - 2\sqrt{s}(E_{\mu_f^+} + E_{\mu_f^-})},
\end{eqnarray}
which provides a convenient way to estimate the missing invariant mass using only the energies and invariant mass of the observed forward muons.

%%%  Kinematical distribution plots (ER 10)  %%%
%%%

The MIM distribution for the signal exhibits a broad peak structure, whose shape and extent depend on the mass of the inert scalars and the available phase space for the invisible final states. The signal involves two nearly degenerate dark-sector particles, resulting in the MIM distribution peaking around twice the dark matter mass.  As shown in \autoref{fig:kine_dist_ER1}(b), for the $\delta_{\rm res}=1\%$ case, the signal distribution rises around 1200 GeV and falls gradually, whereas the most dominant SM background ($\mu_f^+ \mu_f^- \nu \bar{\nu} $) tends to peak at much lower MIM values, typically below 1000 GeV, before quickly falling off. This provides a useful handle for discriminating signal from backgrounds. However, when the energy resolution deteriorates to $\delta_{\rm res}=10\%$, as seen in \autoref{fig:kine_dist_ER10}(b), the signal distribution becomes broader and shifts towards higher MIM values due to energy smearing, while the background distributions remain comparatively similar. This degradation makes it increasingly difficult to apply sharp selection cuts and reduces the effectiveness of MIM as a discriminating observable under poor resolution.

A complementary observable is the total transverse momentum of the di-muon system, $p_\perp^{\mu_f\mu_f} = (p_{\mu_f^+}+ p_{\mu_f^-})_\perp$, which provides an estimate of the missing transverse momentum recoiling against the invisible final state. As shown in \autoref{fig:kine_dist_ER1}(c), the signal exhibits a relatively broad distribution, gradually falling from high values, while the dominant SM backgrounds, particularly $\mu_f^+ \mu_f^- \nu \bar{\nu}$ and $\mu_f^+ \mu_f^- W^+ W^-$, are concentrated at lower $p_\perp^{\mu_f\mu_f}$ values. The signal retains a significant number of events above 200 GeV, extending up to around 1000 GeV, whereas the backgrounds rapidly diminish in this region. This makes $p_\perp^{\mu_f\mu_f}$ an efficient variable for suppressing soft-recoil dominated backgrounds. Moreover, since the shape of this distribution remains fairly stable even under poorer energy resolutions, a hard lower cut around 180-200 GeV can considerably improve the signal-to-background ratio without introducing significant sensitivity to resolution effects.

Next, the pseudorapidity distribution of the first $p_T$ ordered forward muon, $\eta_{\mu_{f_1}}$, shown in \autoref{fig:kine_dist_ER1}(d), confirms that both the signal and background events predominantly populate the extreme forward regions, consistent with the expected VBF topology. The VBF-produced muons typically emerge with large pseudorapidity separations, $\Delta \eta_{\mu_f\mu_f}$, as illustrated in \autoref{fig:kine_dist_ER1}(e). Both signal and background events exhibit large values of $\Delta \eta_{\mu_f\mu_f}$, with distributions populated around $\Delta \eta_{\mu_f\mu_f} \approx 8$, indicating their presence in the forward muon detector acceptance. The azimuthal angle separation between the two forward muons, $\Delta \phi_{\mu_f\mu_f}$, shown in \autoref{fig:kine_dist_ER1}(f), provides complementary information about the angular configuration of the events. While the signal distribution features a distinct dip around $\Delta \phi_{\mu_f\mu_f} = 1.5$, this pattern is absent in the background distributions.

Finally, the minimum and maximum energies of the two forward muons in each event, denoted as $E^{\min}_{\mu_f}$ and $E^{\max}_{\mu_f}$ respectively, are shown in \autoref{fig:kine_dist_ER1}(g) and (h) for the $\delta_{\rm res} = 1\%$ scenario, and in \autoref{fig:kine_dist_ER10}(c) and (d) for the $\delta_{\rm res} = 10\%$ case. These are constructed by identifying the two forward-tagged muons per event and then selecting the smaller and larger of their energy values as $E^{\min}_{\mu_f}$ and $E^{\max}_{\mu_f}$, respectively. These distributions further aid in discriminating the signal from the backgrounds. For $\delta_{\rm res} = 1\%$, the minimum energy of the signal muons sharply falls at around 4.5 TeV, indicating that to produce heavy inert scalars with mass $\sim 550$ GeV, the minimum energy of the final-state muons cannot exceed this value for a 10 TeV MuC, when the energy resolution is good. In contrast, for $\delta_{\rm res} = 10\%$, the $E^{\min}_{\mu_f}$ distribution extends beyond 5 TeV, overlapping significantly with the background, thereby degrading the signal-background discrimination due to the smearing effect. A similar trend is observed for $E^{\max}_{\mu_f}$: at $\delta_{\rm res} = 1\%$, the distribution dies down within 5 TeV, while for $\delta_{\rm res} = 10\%$, it extends beyond 7 TeV, making it difficult to extract any information about the mass of the unknown inert BSM particle. This analysis clearly demonstrates the importance of achieving good beam energy resolution for reliable new physics searches in such compressed mass spectrum scenarios.

%%%  Signal-Background Cut-based analysis (ER 1) %%%
\begin{table}[t]
	\centering
	\hspace*{-1.0cm}
	\renewcommand{\arraystretch}{1.4}
	\begin{tabular}{|l|c|c|c c c c}
		\hline
		\multirow{2}{*}{Cut flow} & \multicolumn{2}{c|}{Signal} & \multicolumn{4}{c|}{Background} \\
		\cline{2-7}
		& BP1 & BP2 & \multicolumn{1}{c|}{$\mu_f^+ \mu_f^- \nu \bar{\nu}$} & \multicolumn{1}{c|}{$\mu_f^+ \mu_f^- W^+ W^-$} & \multicolumn{1}{c|}{$\mu_f^+ \mu_f^- Z Z$} & \multicolumn{1}{c|}{$\mu_f^+ \mu_f^- h$} \\
		\hline\hline
		$\mathcal{C}_0:$ Baseline cuts & 489.75 & 247.76 & \multicolumn{1}{c|}{$8.13\times 10^5$ }  & \multicolumn{1}{c|}{$1.04\times 10^4$} & \multicolumn{1}{c|}{102.35} & \multicolumn{1}{c|}{3.44 } \\
		\hline
		$\mathcal{C}_1: \mathcal{C}_0 + M_{\mu_f \mu_f}>$ 5.0 TeV & 451.58  & 221.87 & \multicolumn{1}{c|}{$4.34\times 10^5$}  & \multicolumn{1}{c|}{$8.60\times 10^3$} & \multicolumn{1}{c|}{88.63} & \multicolumn{1}{c|}{3.42} \\
		\hline
		$\mathcal{C}_2: \mathcal{C}_1 + |\Delta \eta_{\mu_f \mu_f}| < 10.0$ & 427.25 & 211.85 & \multicolumn{1}{c|}{$4.20\times 10^5$ }  & \multicolumn{1}{c|}{$3.73\times 10^3$ } & \multicolumn{1}{c|}{82.82} & \multicolumn{1}{c|}{2.96} \\
		\hline
		$\mathcal{C}_3: \mathcal{C}_2 + \text{MIM} >$ 1.25 TeV  & 345.01 & 211.01 & \multicolumn{1}{c|}{$7.79\times 10^3$} & \multicolumn{1}{c|}{$1.85\times 10^3$ } & \multicolumn{1}{c|}{7.17 } & \multicolumn{1}{c|}{$-$} \\
		\hline	
		$\mathcal{C}_4: \mathcal{C}_3 + E_{\mu_f}^{\text{min}} >$ 2.0 TeV & 308.21 & 185.80 & \multicolumn{1}{c|}{$2.84\times 10^3$ } & \multicolumn{1}{c|}{$1.61\times 10^3$ } & \multicolumn{1}{c|}{6.18 } & \multicolumn{1}{c|}{$-$} \\
		\hline	
		$\mathcal{C}_5: \mathcal{C}_4 + E_{\mu_f}^{\text{max}} <$ 4.8 TeV  & 298.31 & 184.49 &  \multicolumn{1}{c|}{$2.25\times 10^3$ } & \multicolumn{1}{c|}{$1.59\times 10^3$ } & \multicolumn{1}{c|}{6.04 } & \multicolumn{1}{c|}{$-$} \\
		\hline	
		$\mathcal{C}_6: \mathcal{C}_5 + p_\perp^{\mu_f \mu_f} >$ 180 GeV & 201.59 & 117.74 & \multicolumn{1}{c|}{$1.16\times 10^3$ }  & \multicolumn{1}{c|}{547.90 } & \multicolumn{1}{c|}{2.84 } & \multicolumn{1}{c|}{$-$} \\
		\hline	
		$\mathcal{C}_7: \mathcal{C}_6 + |\Delta \phi_{\mu_f \mu_f}- 1.5| > 0.2$ & 188.90 & 111.23 & \multicolumn{1}{c|}{$1.00\times 10^3$ } & \multicolumn{1}{c|}{481.31 } & \multicolumn{1}{c|}{2.06} & \multicolumn{1}{c|}{$-$} \\
		\hline \hline
		Significance at $\int \mathcal{L} = 10 \rm{ab}^{-1}$   & 4.90 & 2.88 &  &  &  &  \\
		\cline{1-3}		
	\end{tabular}
	\caption{Sequential cut-flow table for BP1 and BP2 at a 10 TeV muon collider with 10 ab$^{-1}$ luminosity and $\delta_{\rm res}=1\%$. Event yields are shown after each cumulative selection step, with the final row reporting the signal significance ($\mathcal{S}$) for each BP.}
	\label{Tab:SigBG_CBanal1}
\end{table}	
%%%  

To quantify the sensitivity of the cut-based strategy, we compute the signal significance for BP1 and BP2 at a 10 TeV MuC with 10 ab$^{-1}$ luminosity for the $\delta_{\rm res}=1\%$ case. The cut-based event selection is applied sequentially on the signal and backgrounds, and the number of surviving events at each stage is summarized in \autoref{Tab:SigBG_CBanal1}. The cut-flow begins with the baseline selection cuts ($\mathcal{C}_0$) described in \autoref{eq:BasicCut}, followed by successive kinematic selections using the observables discussed above. Each additional requirement progressively improves the signal-to-background ratio by removing background dominated regions of the kinematic phase space. The final row of the table presents the signal significance, computed as $\mathcal{S}= S/\sqrt{S+B}$, for BP1 and BP2 after applying the complete set of optimized cuts.

The final signal significances for BP1 and BP2 at an integrated luminosity of 10 ab$^{-1}$ are summarized in \autoref{tab:cutbased} for three different muon energy resolutions, $\delta_{\rm res}=1\%$, $5\%$, and $10\%$. As expected, the sensitivity degrades significantly as the energy resolution worsens, with the signal significance for BP1 dropping from 4.90\,$\sigma$ at 1\% resolution to 1.08\,$\sigma$ at 10\%. The situation is even more challenging for BP2 due to its comparatively lower signal cross-section. These results clearly highlight the importance of achieving good beam energy resolution for such challenging semi-invisible final state searches. In the next subsection, we explore the improvement in sensitivity achievable by employing multivariate techniques based on machine learning.

\begin{table}[t!]
	\centering
	\renewcommand{\arraystretch}{1.4}
	\begin{tabular}{|c|c|c|c|}
		\hline
		Benchmark Point & $\delta_\text{res}$ = 1\% & $\delta_\text{res}$ = 5\% & $\delta_\text{res}$ = 10\% \\
		\hline
		BP1 & 4.90\,$\sigma$ & 2.09\,$\sigma$ & 1.08\,$\sigma$ \\
		\hline
		BP2 & 2.88\,$\sigma$ & 1.17\,$\sigma$ & 0.56\,$\sigma$ \\
		\hline
	\end{tabular}
	\caption{Final signal significances for BP1 and BP2 at a 10 TeV MuC with an integrated luminosity of 10 ab$^{-1}$, computed using the optimized cut-based selection for three different muon energy resolutions. The results correspond to the cumulative application of the selection cuts described in \autoref{Tab:SigBG_CBanal1}.} \label{tab:cutbased}
\end{table}

%======================================================================================
\subsection{Multivariate Analysis using XGBoost}

In the previous section, we presented a cut-based analysis for the VBF production of inert scalar pairs at a 10 TeV MuC. The semi-invisible final state, comprising two forward muons and missing energy, was examined for two benchmark points under different muon beam energy resolution scenarios. The corresponding signal significances at an integrated luminosity of 10 ab$^{-1}$ are summarized in \autoref{tab:cutbased}. While the cut-based approach performs reasonably well under ideal resolution, its effectiveness deteriorates considerably as the resolution worsens. This limitation arises primarily because fixed threshold cuts are applied to a limited set of kinematic variables, without making use of potential correlations among them. In scenarios with a compressed mass spectrum between the BSM scalars, the signal kinematic distributions overlap considerably with those of the dominant backgrounds, particularly with $\mu^+ \mu^- \to \mu_f^+ \mu_f^- \nu \bar{\nu}$. As a consequence, the ability of simple cut-based selections to discriminate between signal and background is significantly reduced, leading to suppressed sensitivity. This motivates the adoption of multivariate analysis (MVA) techniques, which can capture subtle, multidimensional differences in event kinematics and improve the overall discriminating power of the search.

In our analysis, we employ XGBoost (XGB) \cite{Chen_2016}, a gradient-boosted decision tree algorithm, to classify events as signal or background based on multiple high-level kinematic variables. XGB constructs an ensemble of decision trees sequentially, with each successive tree focusing on correcting the misclassifications of its predecessors. This approach is particularly well suited for capturing non-linear dependencies between features and remains robust even in the presence of overlapping signal-background distributions~\footnote{Recent analyses~\cite{Ghosh:2025gue, Ghosh:2023ocz} have shown that two distinct leptoquark models, which provide identical final states at the LHC and total production cross sections, can be effectively distinguished using an XGBoost-based multivariate analysis and polarization-sensitive kinematic observables.}. In the following subsections, we outline the feature selection strategy, training configuration, and the performance improvements achieved using this multivariate technique.

%======================================================================================
\subsubsection{Feature Selection}

To train the XGBoost classifier, we start with a set of high-level kinematic observables constructed from the forward-tagged final state muons. All variables are defined based on the $p_T$-ordered muons in each event. After applying the baseline selection cuts, denoted as $\mathcal{C}_0$, we build feature data frames containing the signal and background events. The complete list of kinematic features considered in the initial training set includes:

\begin{itemize}
	\item $p_\perp^{\mu_f\mu_f}$: Transverse momentum of the di-muon system.
	\item $M_{\mu_f\mu_f}$: Invariant mass of the muon pair.
	\item $\Delta\eta_{\mu_f\mu_f}$: Pseudorapidity separation between the two muons.
	\item $\Delta\phi_{\mu_f\mu_f}$: Azimuthal angle separation between the muons.
	\item $\Delta R_{\mu_f\mu_f}$: Angular separation between two muons in $\eta-\phi$ plane.
	\item $E^{\min}_{\mu_f}$, $E^{\max}_{\mu_f}$: Minimum and maximum energy among the two muons.
	\item $E^{\text{tot}}_{\mu_f}$: Total energy of the di-muon system.
	\item $\text{MET}$: Missing transverse energy
	\item $\text{MIM}$: Missing Invariant Mass 
\end{itemize}

\begin{figure}[t]
	\begin{center}
		%\hspace*{-1.cm}
		\includegraphics[width=0.65\linewidth,angle=0]{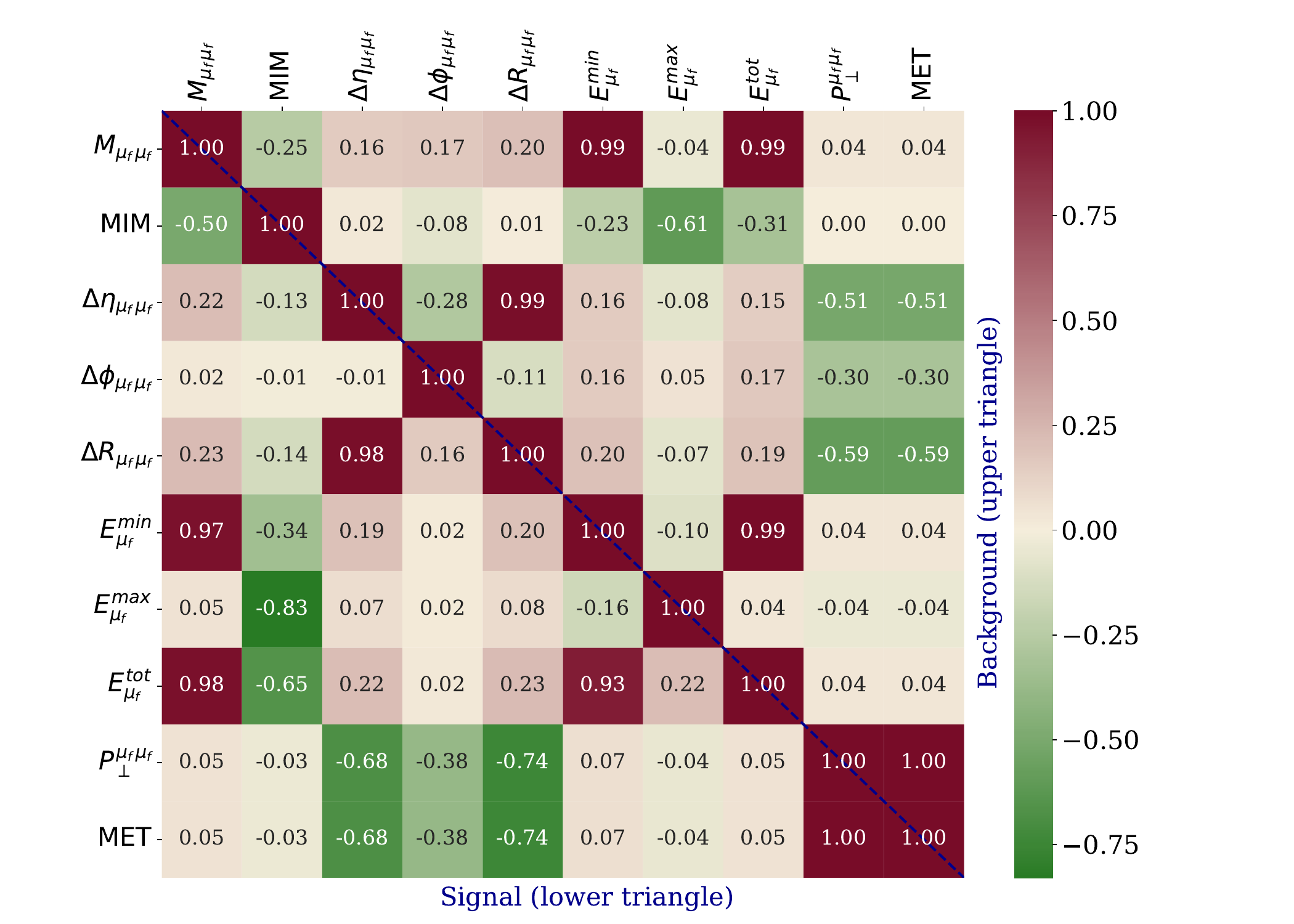}
		\caption{Correlation matrix of the input features for the BP1 scenario with $\delta_{\rm res} = 1\%$. The lower triangle shows the correlations for the signal events, while the upper triangle represents those for the backgrounds. Positive and negative values indicate correlation and anti-correlation, respectively, between feature variables and also visually indicated with color map. Strongly correlated observables are pruned to improve classification performance.} \label{fig:corr_matrix}
	\end{center}
\end{figure}

To improve training efficiency and reduce redundancy, we perform a correlation analysis among the selected input features. Strongly correlated variables often carry overlapping information, which can affect model generalization and reduce the effectiveness of multivariate classifiers. The resulting correlation matrix for the BP1 scenario with $\delta_{\rm res} = 1\%$ is shown in \autoref{fig:corr_matrix}. In this matrix, the diagonal entries correspond to self-correlations and are, by definition, equal to one both for signal and background. With respect to this diagonal entries, the lower triangle represents correlations for the signal events, while the upper triangle shows those for the total background. Positive and negative values indicate correlation and anti-correlation, respectively, between different feature variables and that is also shown with a color map.

%%%  DM Direct Detection  %%%
%%%

As observed in \autoref{fig:corr_matrix}, certain variables exhibit strong correlations, notably between $p_\perp^{\mu_f \mu_f}$ and MET, and between $\Delta \eta_{\mu_f\mu_f}$ and $\Delta R_{\mu_f\mu_f}$. Additionally, we find that $E^{\text{tot}}_{\mu_f}$ is strongly correlated with both $M_{\mu_f\mu_f}$ and $E^{\min}_{\mu_f}$, implying that including all three would introduce redundancy. Since $E^{\min}_{\mu_f}$ and $E^{\max}_{\mu_f}$ together effectively capture the energy distribution of the muon pair, we omit $E^{\text{tot}}_{\mu_f}$ from the final feature set. We have also verified that the correlation matrices for the other benchmark point and different muon beam energy resolution scenarios exhibit a similar pattern. Therefore, for brevity, we present the result for a representative case only. Based on this, we retain only the most discriminating and relatively uncorrelated features for training. Specifically, we drop MET, $\Delta R_{\mu_f\mu_f}$ and $E^{\text{tot}}_{\mu_f}$, as their information is effectively captured by $p_\perp^{\mu_f\mu_f}$, $\Delta \eta_{\mu_f\mu_f}$ and by the pair of $E^{\min}_{\mu_f}$ and $E^{\max}_{\mu_f}$, respectively. This reduces redundancy and improves the classification performance.

With the final set of input features selected, we turn to assigning event weights proportional to their fiducial cross-sections, to appropriately reflect the expected event rates during training. The fiducial cross-section for each process is computed as:

\begin{eqnarray}
	\sigma_{\text{fid}} = \frac{\sigma \times n_{\rm obs}}{N_{\text{gen}}}
\end{eqnarray}

where $\sigma$ is the parton-level cross-section obtained from \texttt{MadGraph5\_aMC@NLO}, $n_{\rm obs}$ is the number of events passing the baseline selection cuts $\mathcal{C}_0$, and $N_{\text{gen}}$ is the total number of generated events for that process. The corresponding event weight is then defined as:
\begin{eqnarray}
	w_i = \frac{\sigma_{\text{fid}}^{(i)}} {\sum_{j}\sigma_{\text{fid}}^{(j)}},
\end{eqnarray}
ensuring that the total sum of event weights is normalized to unity. This weighting scheme maintains numerical stability during training and avoids bias, particularly important when the signal and background cross-sections differ by several orders of magnitude (in our case, by approximately five orders).

In addition to weighting, we address class imbalance arising from the differing numbers of generated events for signal and background. Since the total number of background events generated by \texttt{MadGraph5\_aMC@NLO} exceeds that of the signal by a large factor, we mitigate class imbalance by applying a resampling strategy. Specifically, we downsample the background events to bring their count closer to that of the signal ($\sim 5 \times 10^6 $), while preserving the shapes of their kinematic distributions. This ensures balanced and unbiased training of the classifier.

%======================================================================================
\subsubsection{Training Setup and Classifier Performance}

In this subsection, we describe the training procedure and classifier performance of the XGBoost model for BP1 with a muon energy resolution of $\delta_{\rm res} = 1\%$. This configuration offers the clearest separation between signal and background and is treated as the baseline case for our analysis. Additionally, we present the corresponding classifier performance, including signal-background efficiency curves, for the same benchmark point at $\delta_{\rm res} = 5\%$ and $10\%$, using an identical training setup.

The combined signal and background dataset is randomly split into training and test sets using a 60:40 ratio. As discussed in the previous section, the training set is balanced such that it contains equal numbers of signal and background events, and the total training weight for each class is normalized to unity. This ensures that the classifier learns to discriminate between signal and background without bias toward the more abundant class. Furthermore, 30\% of the training set is reserved as a validation set to monitor classifier performance during training and to guard against overfitting. 

The classifier is implemented using the \texttt{XGBClassifier} from the \texttt{xgboost} Python package. The training is performed using a binary logistic loss function, with the hyperparameters set as follows:

\begin{verbatim}
	xgb_clf = XGBClassifier(objective='binary:logistic', booster='gbtree', 
	learning_rate=0.082, max_depth=5, n_estimators=800, gamma=0, subsample=0.52,
	colsample_bytree=0.6, colsample_bylevel=0.5, min_child_weight=4, reg_lambda=1, 
	reg_alpha=4, scale_pos_weight=1, eval_metric=['auc', 'logloss'])
\end{verbatim}

The hyperparameters are chosen to balance model flexibility and generalization performance. A relatively small \texttt{learning\_rate} and moderate number of \texttt{n\_estimators} prevent overfitting while maintaining model capacity. The \texttt{subsample} and \texttt{colsample\_bytree} parameters introduce controlled randomness, improving robustness by decorrelating trees. The regularization parameter \texttt{min\_child\_weight} is set to 4 to reduce sensitivity to statistical fluctuations in the low event count regions typical of our final event selections. These values were manually tuned to maximize the area under the receiver operating characteristic (ROC) curve and minimize the binary logistic loss (log loss) on the validation set, ensuring both classification power and numerical stability.

%%%  Sig_BG efficiency with AUC  %%%

\begin{figure}[t]
	\begin{center}
		\hspace*{-1.5cm}
		\mbox{\subfigure[]{\includegraphics[width=0.38\linewidth,angle=0]{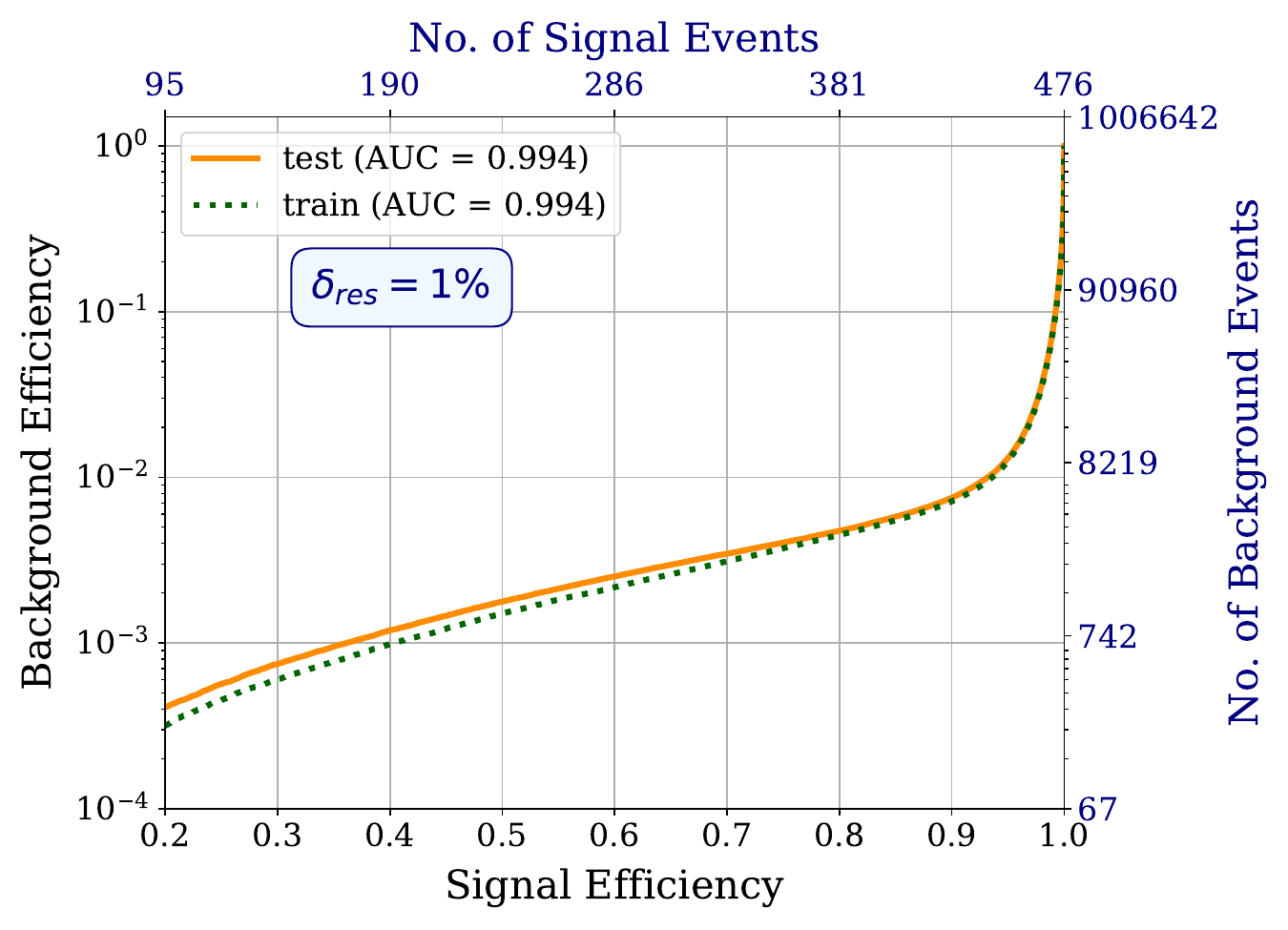}}\quad 
			\subfigure[]{\includegraphics[width=0.38\linewidth,angle=0]{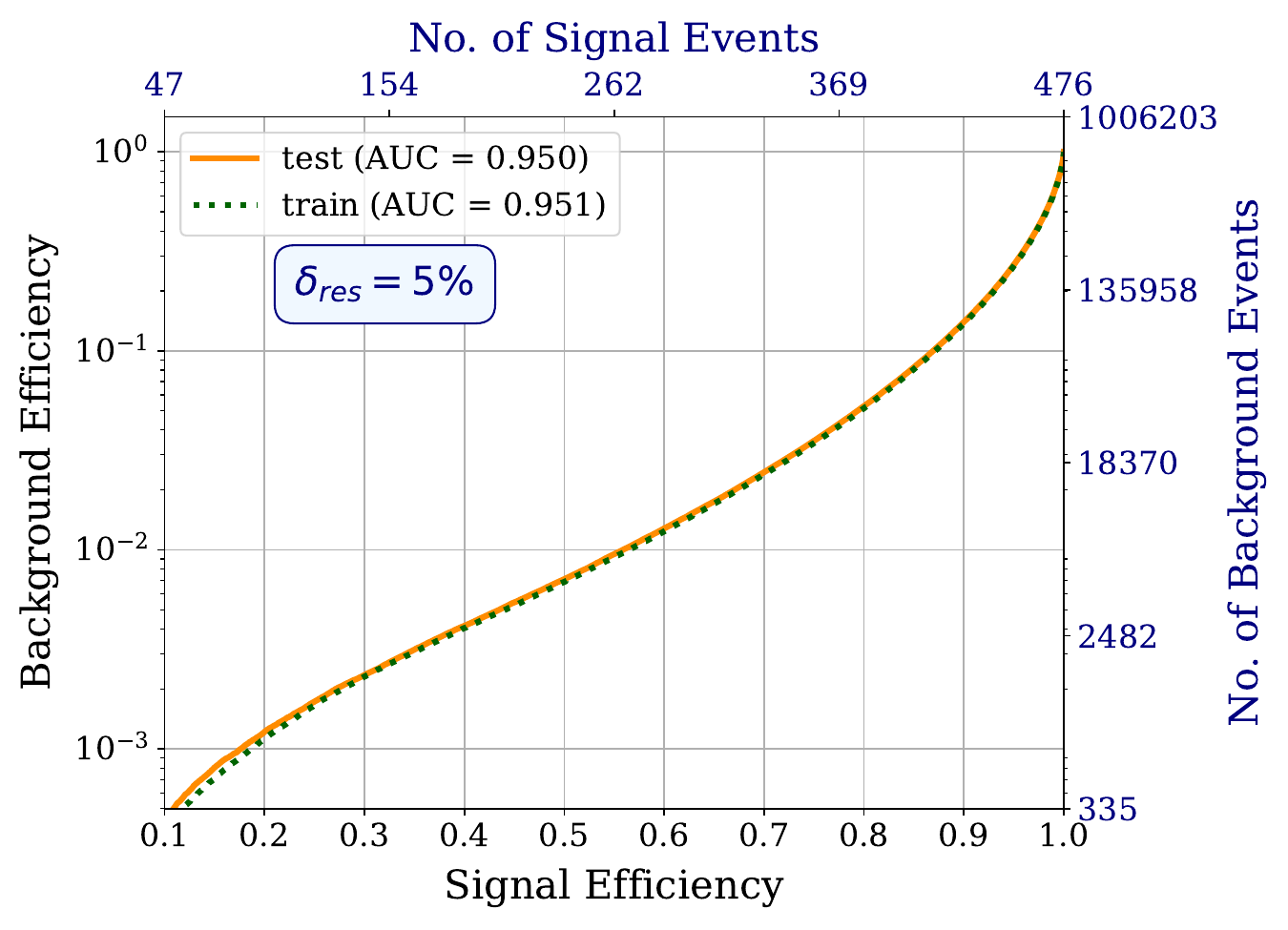}}\quad 
			\subfigure[]{\includegraphics[width=0.38\linewidth,angle=0]{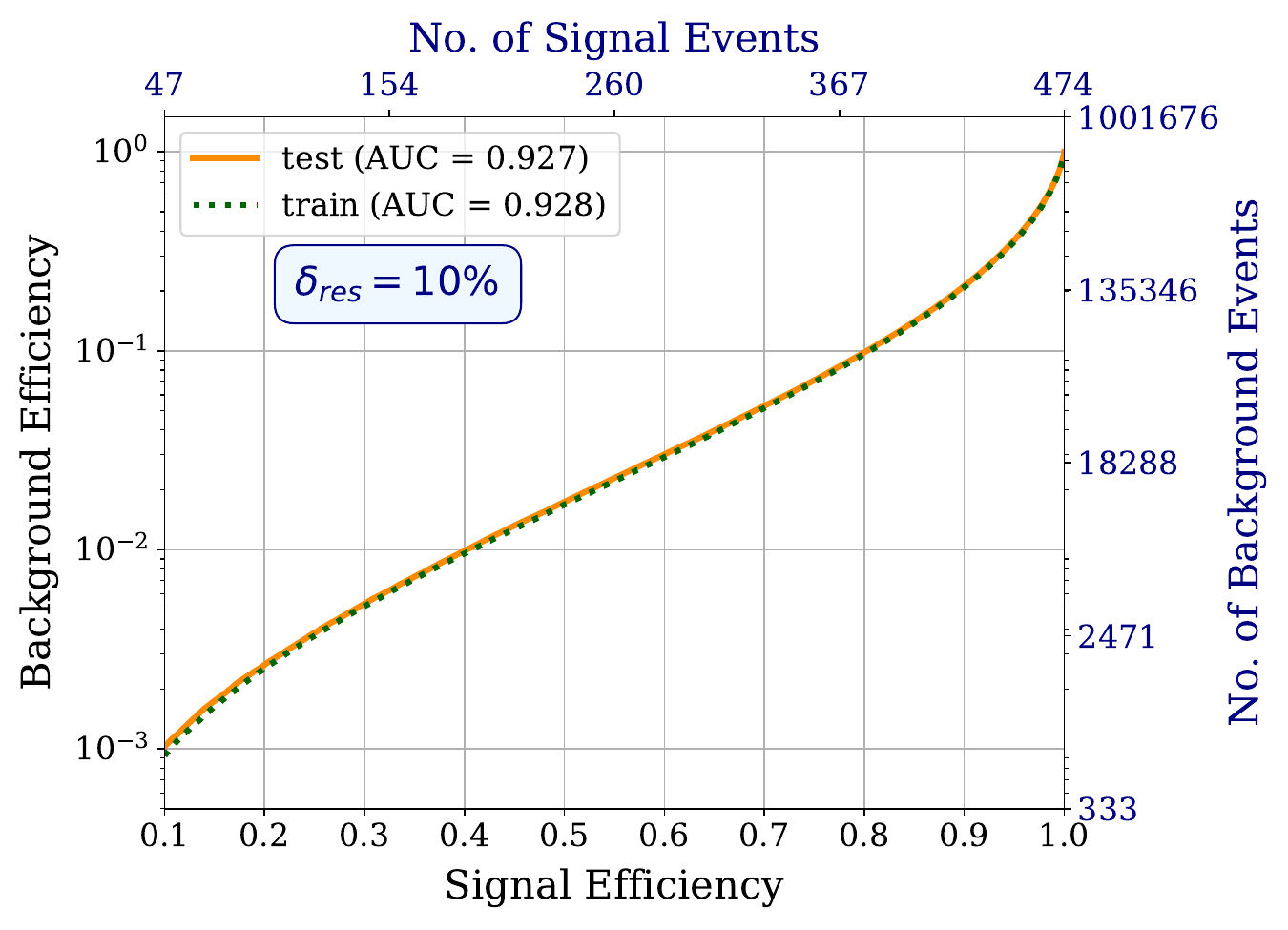}}}
		\caption{ROC curves for both training and test datasets for BP1, corresponding to muon energy resolutions of $\delta_{\rm res} = 1\%$, $5\%$, and $10\%$ in (a), (b), and (c), respectively. The number of expected signal and background events for a luminosity of 10~ab$^{-1}$ at a 10 TeV MuC is shown on the top and right sides of each plot.} \label{fig:efficiency_AUC}
	\end{center}
\end{figure}
%%%

The classifier performance is evaluated using ROC curves for both the training and test datasets. \autoref{fig:efficiency_AUC} shows the ROC curves obtained for BP1 with muon energy resolutions of $\delta_{\rm res} = 1\%$, $5\%$, and $10\%$ in panels (a), (b), and (c), respectively. The test and training ROC curves are shown as orange solid and green dotted lines, respectively. The area under the curve (AUC) is quoted for both datasets in each case. As expected, the classifier achieves excellent separation for the $\delta_{\rm res} = 1\%$ scenario, with an AUC value of 0.994 for both training and testing datasets. The performance degrades moderately as the energy resolution worsens, with AUC values reducing to 0.950 and 0.927 for $\delta_{\rm res} = 5\%$ and $10\%$, respectively. This behavior reflects the increasing overlap in kinematic distributions between the signal and background at poorer resolutions, as discussed in earlier sections. The number of expected signal and background events corresponding to each signal efficiency point is indicated on the top and right axes of the plots, providing a direct connection between classifier performance and event yields at an integrated luminosity of 10~ab$^{-1}$ at a 10 TeV MuC.

\begin{figure}[t]
	\begin{center}
		\hspace*{-1.5cm}
		\mbox{\subfigure[]{\includegraphics[width=0.38\linewidth,angle=0]{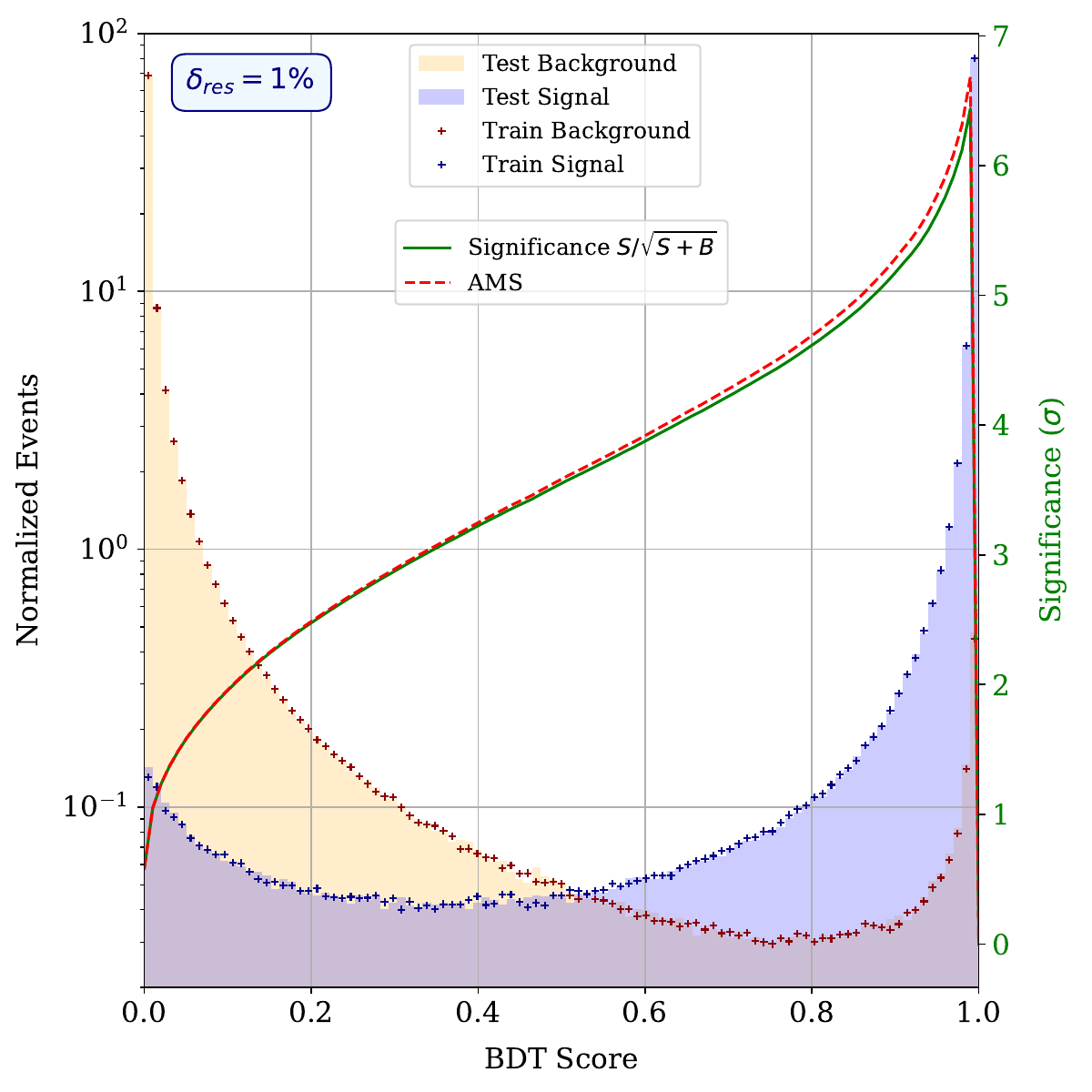}}\quad 
			\subfigure[]{\includegraphics[width=0.38\linewidth,angle=0]{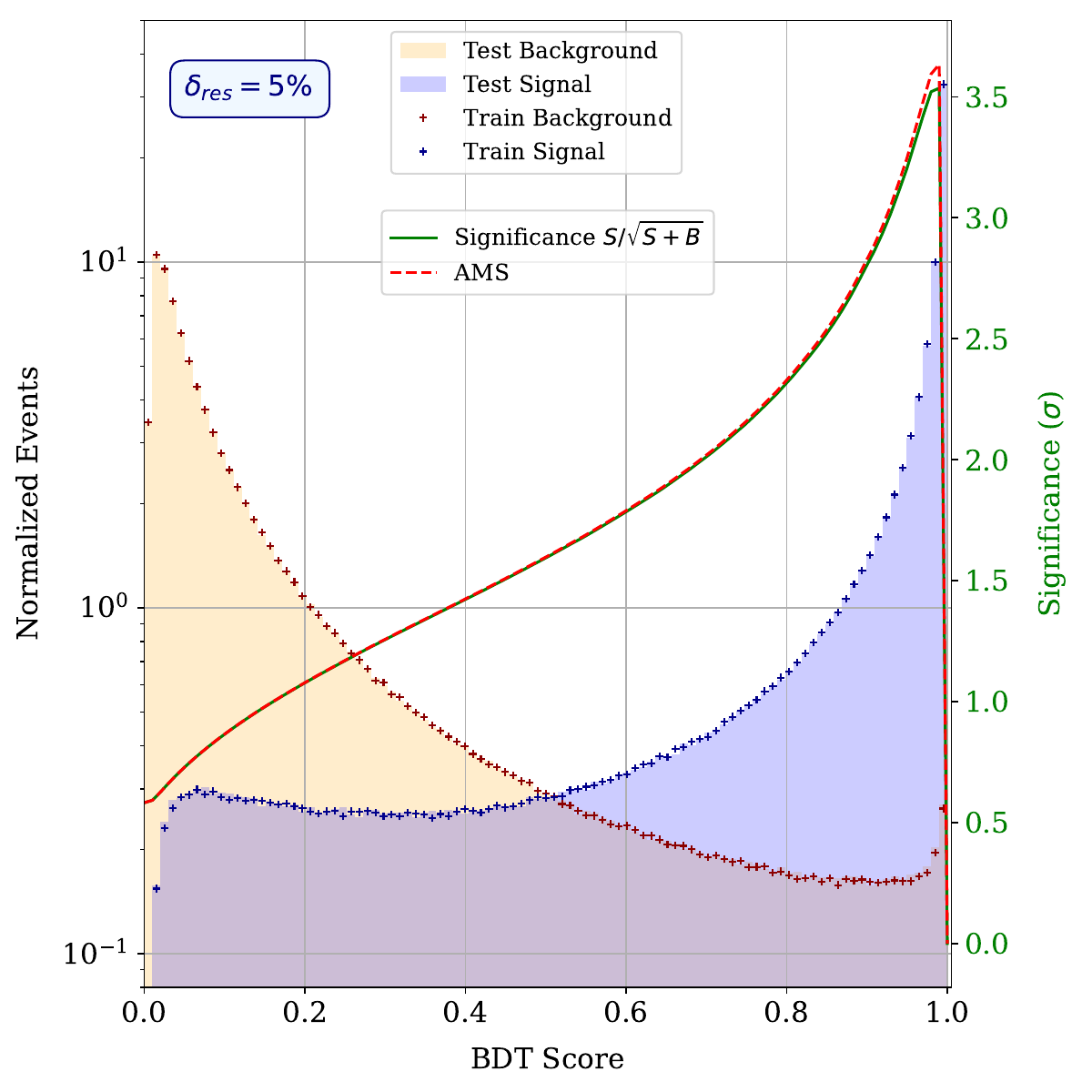}}\quad 
			\subfigure[]{\includegraphics[width=0.38\linewidth,angle=0]{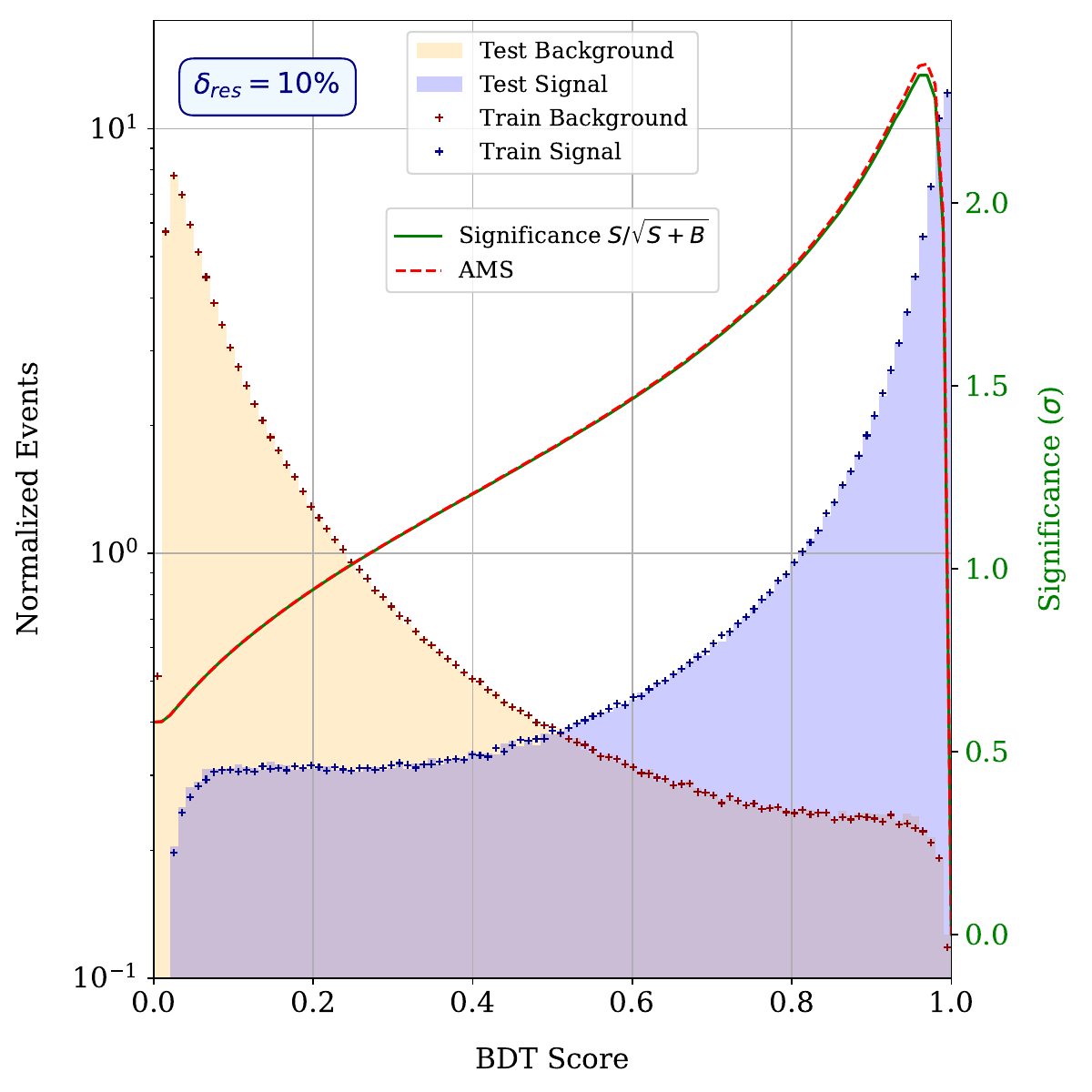}}}
		\caption{XGBoost classifier output scores for BP1 at muon energy resolutions of $\delta_{\rm res} = 1\%$, $5\%$, and $10\%$ shown in panels (a), (b), and (c), respectively. The signal and background events are shown separately for the training and test datasets, with signal events in dark blue (train) and light blue (test), and background events in red (train) and light orange (test). The solid green curve illustrates the significance computed using $\mathcal{S} =S/\sqrt{S+B}$, while the red dashed curve represents the significance computed using AMS. Both significance curves are plotted as functions of the classifier score threshold. The secondary y-axis on the right shows the significance values, marked with green ticks. }\label{fig:significance}
	\end{center}
\end{figure}

Following the ROC-based performance evaluation, we examine the classifier score distributions for signal and background events to determine the optimal selection threshold that maximizes the statistical significance. \autoref{fig:significance} shows the normalized XGBoost classifier output score distributions for BP1, corresponding to muon energy resolutions of $\delta_{\rm res} = 1\%$, $5\%$, and $10\%$. Signal events are displayed in dark blue (train) and light blue (test), while background events appear in red (train) and light orange (test).

Overlaid on these distributions are the expected significance curves, plotted on the secondary y-axis (right). The solid green curve represents the standard significance metric, $\mathcal{S}= S/\sqrt{S + B}$, while the dashed red curve corresponds to the Approximate Median Significance (AMS), defined as:
\begin{eqnarray}
	\text{AMS} = \sqrt{2 \left[ (S + B) \ln\left(1 + \frac{S}{B} \right) - S \right]},
\end{eqnarray}
where $S$ and $B$ denote the expected signal and background event counts above a given XGB score threshold.

%%%  Signal significance  %%%
%%%

As anticipated, the signal distribution peaks sharply at score values close to One, while background events are concentrated predominantly near Zero, demonstrating strong separation between signal and background. The significance curves exhibit a clear maximum in the high-score region, identifying the optimal classifier threshold for event selection. For BP1, the maximum AMS significance reaches approximately $6.9\,\sigma$, $3.7\,\sigma$, and $2.4\,\sigma$ for $\delta_{\rm res}=1\%$, $5\%$, and $10\%$, respectively, evaluated at an integrated luminosity of 10 ab$^{-1}$. These results represent a substantial improvement over the corresponding cut-based analysis at the same energy resolutions, as summarized in \autoref{tab:cutbased}.

%%%  Optimal Working Points Table %%%
\begin{table}[t]
	\centering
	%\hspace*{-1.5cm}
	\renewcommand{\arraystretch}{1.4}
	\begin{tabular}{|c|c|c|c|c|c|c|}
		\hline
		\multirow{2.2}{*}{$\delta_{\rm res}$ }  & \multicolumn{2}{c|}{Signal ($\epsilon_S$)} & \multicolumn{2}{c|}{Background ($\epsilon_B \times 10^{3} $)} & \multicolumn{2}{c|}{\makecell{Signal Significance \\ $\mathcal{S}$ (AMS)}} \\
		\cline{2-7}
		 in \% & BP1 & BP2 & BP1 & BP2 & BP1 & BP2 \\
		\hline\hline
		 1\% & 387.4 (0.81) & 219.0 (0.91)  & 3233.9 ($5.27$)  & 3054.8 ($4.57$)  & 6.60 (6.86)  & 3.83 (3.92) \\
		\hline
		 5\% & 157.6 (0.33) & 110.6 (0.46) & 1829.4 ($3.46$)  & 2158.6 ($3.23$) & 3.57 (3.67) & 2.32 (2.36) \\
		\hline 
		10\% & 143.7 (0.30) & 68.7 (0.28)  & 3596.0 ($5.38$)  & 2399.1 ($3.40$) & 2.35 (2.38) & 1.38 (1.40) \\
		\hline
	\end{tabular}
	\caption{Signal and background yields after applying the optimal XGBoost classifier thresholds at different muon energy resolutions for BP1 and BP2. Values in parentheses indicate the selection efficiencies $\epsilon_S$, $\epsilon_B$, and the AMS significance, respectively.}
	\label{Tab:xgb_results}
\end{table}	
%%%  
%%%%  Optimal Working Points Table %%%
%\begin{table}[t]
%	\centering
%	\hspace*{-1.5cm}
%	\renewcommand{\arraystretch}{1.2}
%	\begin{tabular}{|c|c|c|c|c|c|c|}
%		\hline
%		\multirow{2.2}{*}{\makecell{$\delta_{\rm res}$ \\ (in \%)}}  & \multicolumn{2}{c|}{\makecell{Signal \\ ($\epsilon_S$)}} & \multicolumn{2}{c|}{\makecell{Background \\ ($\epsilon_B$)}} & \multicolumn{2}{c|}{\makecell{Signal Significance \\ $\mathcal{S}$ (AMS)}} \\
%		\cline{2-7}
%		  & BP1 & BP2 & BP1 & BP2 & BP1 & BP2 \\
%		\hline\hline
%		 1\% & 387.4 & 219.0   & 3233.9   & 3054.8  & 6.60 (6.86)  & 3.83 (3.92) \\
%		 	& (0.81) & (0.91)  & ($5.27 \times 10^{-3}$)  & ($4.57 \times 10^{-3}$)  &   &  \\
%		\hline
%		 5\% & 157.6 & 110.6 & 1829.4  & 2158.6  & 3.57 (3.67) & 2.32 (2.36) \\
%		  	& (0.33) & (0.46) & ($3.46 \times 10^{-3}$)  & ($3.23 \times 10^{-3}$) &  &  \\
%		\hline 
%		10\% & 143.7 & 68.7 & 3596.0   & 2399.1 & 2.35 (2.38) & 1.38 (1.40) \\
%			 & (0.30) & (0.28)  &($5.38 \times 10^{-3}$)  & ($3.40 \times 10^{-3}$) &  &  \\
%		\hline
%	\end{tabular}
%	\caption{Signal and background yields after applying the optimal XGBoost classifier thresholds at different muon energy resolutions for BP1 and BP2. Values in parentheses indicate the selection efficiencies $\epsilon_S$, $\epsilon_B$, and the AMS significance, respectively.}
%	\label{Tab:xgb_results}
%\end{table}	
%%%%  

To comprehensively summarize the classifier performance, \autoref{Tab:xgb_results} presents the signal and background yields after applying the optimal XGBoost classifier thresholds for both BP1 and BP2 across the three muon energy resolution scenarios. In addition to the expected event yields, the table also lists the corresponding selection efficiencies for signal ($\epsilon_S$) and background ($\epsilon_B$), as well as the achieved signal significance values computed using both the standard $S/\sqrt{S+B}$ and the AMS formula. While the preceding discussion focused on BP1, this table includes, for completeness, the results for BP2, which exhibits similar qualitative behavior though with overall reduced significance due to the smaller signal cross-section.

As expected, the signal significance decreases with worsening energy resolution for both BPs, reflecting the reduced separation power at lower detector precision. Notably, for BP1 at $\delta_{\rm res}=1\%$, a significance of 6.60\,$\sigma$ ($6.8\,\sigma$ using AMS) is achieved, while BP2 yields 3.83\,$\sigma$ (3.92\,$\sigma$) under the same conditions. Even at 10\% resolution, the classifier retains meaningful discriminating power, demonstrating the robustness of the multivariate approach in probing compressed mass spectra in inert scalar models, even in difficult final states with limited visible activity.

%======================================================================================
\section{Conclusion}
\label{sec:conclusion}

In this work, we investigate the prospects for probing the Inert scalar Doublet Model in a compressed dark sector mass spectrum scenario at a future 10 TeV Muon Collider (MuC). Focusing on the vector boson fusion production of inert scalar pairs, we consider the challenging final state containing only two forward muons and no visible activity in the central detector. In such compressed scenarios, large missing energy signatures are absent as the pair-produced inert scalars tend to balance each other’s transverse momentum. At hadron colliders, the difficulty of cleanly tagging exactly two forward jets in the presence of a large QCD radiation, ISR/FSR, and pile-up backgrounds makes this semi-invisible signature particularly difficult to observe. In contrast, a clean high energy lepton collider like the MuC not only offers a significantly enhanced VBF cross-section for a few hundred GeV electroweak states, but also provides forward muon detection capabilities and decent background control, thereby enabling sensitive searches for such elusive signals.

The benchmark points used in our collider analysis are selected based on a detailed dark matter phenomenology study of the IDM. Constraints from relic density measurements, direct detection limits from recent LZ experiment, and the neutrino floor are carefully considered to identify viable regions of parameter space. The two benchmark points represent characteristic scenarios within the narrow allowed region, ensuring both phenomenological consistency and collider testability.

A detailed cut-based analysis is performed using several high-level kinematic observables constructed from the two forward-tagged muons, including the invariant mass of the di-muon system, missing invariant mass, total transverse momentum, pseudorapidity separation, azimuthal angle separation, and individual muon energy distributions. The impact of muon beam energy resolution is carefully studied by considering three benchmark values: $\delta_{\rm res} = 1\%, 5\%$, and $10\%$. The analysis demonstrates that the achievable signal significance strongly depends on the beam energy resolution, with significances dropping from 4.90$\sigma$ to 1.08$\sigma$ for BP1 as the resolution deteriorates from $1\%$ to $10\%$.

To improve sensitivity in regions where the cut-based analysis loses discrimination power, particularly under poorer energy resolution, we implement a multivariate analysis using the XGBoost machine learning classifier. The multivariate approach achieve a visible improvement in performance by capturing correlations among kinematic variables, leading to enhanced signal significance compared to the cut-based method. Notably, for the $\delta_{\rm res}=10\%$ case, where the cut-based significance for BP1 was limited to 1.08$\sigma$, the ML-based analysis improves this to 2.35$\sigma$. This result highlights the robustness and utility of multivariate techniques in probing compressed electroweak scenarios at future colliders, even under challenging detector performance conditions.

In summary, our study demonstrates that a high energy Muon Collider, equipped with efficient forward muon detection and good energy resolution, offers a uniquely clean and powerful platform to probe electroweak sectors with compressed mass spectra, which have long remained experimentally challenging for hadron colliders.

%======================================================================================

%======================================================================================
\section*{Acknowledgements}
The research work at the Physical Research Laboratory (PRL) is funded by the Department of Space, Government of India.
The authors gratefully acknowledge the resources provided by the Param Vikram-1000 High Performance Computing Cluster and the TDP project infrastructure at PRL, which were extensively used for various computational aspects of this work. Authors also acknowledge Deepanshu Srivastava and Snehashis Parashar for useful discussions regarding the machine learning analysis.

\section*{Data Availability Statement}
Simulated data are generated with standard HEP tools for this study. Full details of the simulation procedures are included in the article.

%\newpage
%======================================================================================
\appendix    
%======================================================================================

\bibliographystyle{JHEP}
\bibliography{ref_IDM_MuC}
%======================================================================================
\end{document}